\documentclass[journal]{IEEEtran}
\usepackage{rotating}
\usepackage{amsmath}
\usepackage{graphicx}%add figures
\usepackage{subfigure}
\usepackage{algpseudocode}
\usepackage{algorithm}
\usepackage{algorithmicx}
\usepackage{algpseudocode}
\usepackage{amssymb}
\usepackage{float}
 \usepackage{verbatim}
\usepackage{multirow}
\usepackage{makecell}
\usepackage{hyperref}
\usepackage{graphics} 
\usepackage{diagbox}
\usepackage{color}
\usepackage{multirow,booktabs}
\usepackage{adjustbox}

\usepackage[table]{xcolor} % 引入xcolor包并使用table选项
\usepackage{cleveref}

\usepackage[numbers,sort&compress]{natbib}%用于引用编号压缩
\begin{document}

\title{DS-Pnet: FM-Based Positioning via Downsampling}

\author{Shilian~Zheng,
        Xinjiang~Qiu,
        Luxin~Zhang,
        Quan~Lin,
        Zhijin~Zhao,
        and Xiaoniu~Yang
        % <-this % stops a space

% \thanks{This work was supported in part by the National Basic Scientific Research of China under Grant JCKY2023110C099. \textit{(Corresponding author: Shilian Zheng and Xiaoniu Yang.)}}% <-this % stops a space

\thanks{S. Zheng is with the College of Communication Engineering, Hangzhou Dianzi University, Hangzhou 310018, China. He is also National Key Laboratory of Electromagnetic Space Security,
Jiaxing 314033, China (e-mail: lianshizheng@126.com).}

\thanks{X. Qiu, Q. Lin, and Z. Zhao are with the College of Communication Engineering, Hangzhou Dianzi University, Hangzhou 310018, China (e-mail: 232080090@hdu.edu.cn; 232080162@hdu.edu.cn; zhaozj03@hdu.edu.cn).}

\thanks{L. Zhang and X. Yang are with {{National Key Laboratory of Electromagnetic Space Security}},
Jiaxing 314033, China (e-mail: lxzhangMr@126.com; 
yxn2117@126.com).}% <-this % stops a space
% <-this % stops a space

}% <-this % stops a space

\maketitle

\begin{abstract}
%Frequency Modulation (FM) broadcast signals, regarded as opportunistic signals, hold significant potential for indoor and outdoor positioning applications. The existing FM-based positioning methods primarily rely on Received Signal Strength (RSS) for positioning, the accuracy of which needs improvement. In this paper, we introduce FM-Pnet, an end-to-end FM-based positioning method that leverages deep learning. This method utilizes the time-frequency representation of FM signals as network input, enabling automatically learning of deep features for positioning. We also propose two strategies, noise injection and enriching training samples, to enhance the model's generalization performance over long time spans. We construct datasets for both indoor and outdoor scenarios and conduct extensive experiments to validate the performance of our proposed method. Experimental results demonstrate that FM-Pnet significantly outperforms traditional RSS-based positioning methods in terms of both positioning accuracy and stability.
In this paper we present DS-Pnet, a novel framework for FM signal-based positioning that addresses the challenges of high computational complexity and limited deployment in resource-constrained environments. Two downsampling methods—IQ signal downsampling and time-frequency representation downsampling—are proposed to reduce data dimensionality while preserving critical positioning features. By integrating with the lightweight MobileViT-XS neural network, the framework achieves high positioning accuracy with significantly reduced computational demands. Experiments on real-world FM signal datasets demonstrate that DS-Pnet achieves superior performance in both indoor and outdoor scenarios, with space and time complexity reductions of approximately 87\% and 99.5\%, respectively, compared to an existing method, FM-Pnet. Despite the high compression, DS-Pnet maintains robust positioning accuracy, offering an optimal balance between efficiency and precision.
\end{abstract}


\section{Introduction}
\IEEEPARstart{I}{n} recent years, with the growing demand for location-based services such as autonomous driving  \cite{1}, the Internet of Vehicles (IoV) \cite{2}, and the Internet of Things (IoT) \cite{3}, there has been an increasing need for high-precision, low-latency indoor and outdoor positioning technologies. In open outdoor environments, systems like satellite-based navigation, such as the Global Positioning System (GPS) \cite{4} or the BeiDou Navigation Satellite System (BNSS) \cite{5}  can provide positioning accuracy at the meter to decimeter level without relying on ground infrastructure \cite{6}. When supplemented by ground-based facilities, this accuracy can be further enhanced to the centimeter level \cite{7}. However, in environments with dense buildings (such as commercial centers and underground parking lots), positioning reliability and accuracy are significantly reduced due to severe blockage and attenuation of satellite signals \cite{8}. Against this backdrop, the development of positioning technologies that operate independently of existing satellite systems has garnered increasing attention. 

Currently, opportunistic signals such as analog modulated broadcast signals \cite{9}, Bluetooth signals \cite{10}, WiFi signals \cite{11}, ultra-wideband (UWB) signals \cite{12}, and visible light signals \cite{13} are widely used in positioning research. Among these, WiFi-based fingerprinting positioning schemes have attracted particular attention due to the ubiquity and portability of WiFi devices. WiFi fingerprinting positioning estimates location by matching collected wireless signal features with a pre-established fingerprint database. In WiFi fingerprinting, two types of features are commonly used as fingerprints: received signal strength (RSS) \cite{14} and channel state information (CSI) \cite{15}. Compared to RSS, CSI belongs to the physical layer's fine-grained information and can provide rich multipath effect features, thus significantly improving positioning accuracy. In recent years, researchers have proposed deep learning-based methods that learn CSI features from WiFi signals, further enhancing the accuracy of indoor positioning \cite{16}, \cite{17}. However, WiFi signals have limited coverage and weak penetration, and in practice, they perform better in indoor environments. In contrast, FM signals operate in the 87-108 MHz frequency band, with a lower frequency and reduced sensitivity to environmental changes, offering a wider coverage range and stronger penetration. FM signals can easily cover hundreds of kilometers and have excellent indoor penetration capabilities. Additionally, there is a large number of commercial and amateur FM broadcasts worldwide, which do not require additional infrastructure, making them highly practical. Numerous studies have demonstrated that FM signals exhibit outstanding positioning performance \cite{18}, \cite{19}, \cite{20}. For example, Li et al. \cite{21} proposed an indoor vehicle positioning method based on deep learning and FM fingerprinting, achieving accurate and practical indoor positioning using FM signals. Zheng et al. \cite{22} proposed FM-Pnet, a deep learning-based approach that employs time-frequency representations of FM signals for precise indoor and outdoor positioning. The method demonstrates significant improvements in accuracy and robustness compared to traditional RSS-based techniques across various scenarios.

Although the aforementioned positioning methods based on opportunistic signals achieve excellent positioning accuracy, computational complexity is a crucial factor that cannot be overlooked when deploying deep learning-based positioning applications in real-world scenarios. Existing methods primarily focus on enhancing positioning performance but do not sufficiently evaluate or optimize computational complexity, which limits their applicability in resource-constrained environments. To address this issue, researchers have proposed various methods to reduce computational complexity. In studies utilizing channel state information (CSI) based on WiFi signals for positioning, the high dimensionality and redundancy of CSI data significantly increase computational complexity when directly used for positioning \cite{23}.  To mitigate this, some studies have introduced dimensionality reduction techniques, such as using Principal Component Analysis (PCA) \cite{24}, \cite{25} or deep autoencoders \cite{26} to reduce the dimensionality of CSI data. However, FM signals contain abundant location information due to multipath effects unique to different locations. When wideband FM signals are received at a fixed location with a high sampling rate, the resulting data volume is substantial, and directly processing this data would increase the computational burden. FM broadcast signals generally contain only a few narrowband signals within the sampled frequency band, and these narrowband signals demonstrate sparsity in the time-frequency representation generated by the Short-Time Fourier Transform (STFT). Compared to background noise or irrelevant frequency bands, the narrowband regions enriched with these signals contribute more significantly to the positioning task.

Therefore, we consider that appropriately downsampling FM signals and extracting features to reduce data dimensionality while preserving essential location information is an effective strategy. Based on this idea, this paper proposes a downsampling-based FM signal positioning method named DS-Pnet. This method downsamples the input signal from two distinct perspectives to effectively extract features and simplify computations. The downsampled time-frequency representations are then fed into a deep neural network for location estimation. In the model design phase, to address the requirements of resource-constrained environments or real-time applications, we selected the lightweight MobileViT-XS neural network to reduce both model size and computational complexity. Our contributions are as follows:
\begin{itemize}
% \item We propose a low-complexity FM signal positioning framework called DS-Pnet. DS-Pnet extracts effective features from downsampled time-frequency representations using deep learning, achieving a positioning method that performs well in both indoor and outdoor scenarios.  
\item We propose a more efficient FM-based positioning framework, referred to as DS-Pnet. This framework leverages deep learning to extract features from the downsampled time-frequency representations of FM signals. By achieving a favorable balance between positioning accuracy and computational efficiency, DS-Pnet demonstrates superior performance in both indoor and outdoor environments.
\item We explore multiple downsampling methods, including IQ signal downsampling, time-frequency direct downsampling, and time-frequency attention downsampling, to reduce computational complexity while maintaining the positioning performance of DS-Pnet.

% \item We conduct extensive experiments to evaluate the performance of our proposed DS-Pnet with different downsampling factors and methods. The experimental results show that attention mechanism-based downsampling in DS-Pnet achieves satisfactory performance in both indoor and outdoor scenarios. 
\item We conduct a comprehensive experimental analysis using real FM signal datasets with various downsampling factors and strategies. The results demonstrate that DS-Pnet achieves favorable performance across all downsampling factors, maintaining exceptional stability and accuracy. The attention-based downsampling method is particularly effective, achieving both high efficiency and precision, even at high downsampling rates.

\item We analyze the complexity of the proposed method, including space complexity and time complexity. Compared to FM-Pnet, the space complexity is reduced by approximately 87\%, and the time complexity is reduced by up to 99.5\%, without compromising
accuracy. These findings highlight the potential of DS-Pnet
for scalable and precise positioning solutions, particularly in
resource-constrained environment.
% \item We conduct extensive experiments to evaluate the performance of our proposed FM-Pnet in various scenarios. The experimental results show that the proposed FM-Pnet outperforms three existing RSS-based methods. The performance of FM-Pnet can be further enhanced in the frequency domain or in the time domain by increasing the bandwidth of the collected signals or extending the length of each sample. Additionally, through the two proposed enhancement methods, the generalization of FM-Pnet over a large time span can be improved.
\end{itemize}

The rest of this paper are organized as follows. In Sec. \ref{sec2},  we formulate the problem. In Sec. \ref{sec3}, we introduce our proposed DS-Pnet in detail. In Sec. \ref{sec4}, we compare the positioning performance of DS-Pnet under different downsampling factors and methods.  Finally, we provide the concluding remarks in Sec. \ref{sec5}.

\section{Problem Formulation}
\label{sec2}
 
%  \begin{figure}[t]
%     \centering 
%     \includegraphics[width=7cm]{figs/wentijianmo.eps}  % 8cm
%     \caption{Problem modeling for FM-based Positioning.}
%     \label{fig1}
% \end{figure}

The modeling of the FM signal-based positioning problem can be described as extracting features from the received FM signals to infer the location of the receiving point. Suppose the target is located within a region divided into grid cells, where each grid cell corresponds to a position coordinate. The FM signal received at the target point can be represented as:
\begin{equation}
s(n)= \sum _{i=1}^{K}g_{i}(n)*o_i(n)+ w(n), n=0,1,\dots,L-1,
\label{eq1}
\end{equation}
where $s(n)$ denotes the received signal, $K$  indicates the number of narrowband FM signals in the frequency band, and $L$ represents the signal length. Each narrowband FM signal is expressed in discrete time as $o_{i}(n)$, with $g_{i}(n)$ representing its corresponding channel response. $w(n)$ refers to the noise, which includes additive white Gaussian noise (AWGN) and other random interference. Since the signal transmission paths differ across locations, the features embedded in the received FM signals can be used to distinguish different locations. 

If FM signals can be collected in advance at each location, the positioning task transforms into associating the processed signals with their respective spatial coordinates:
\begin{equation}
s_d(m)=\mathcal{D}\{\mathcal{T}(s(n))\},
\label{eq2}
\end{equation}
\begin{equation}
\mathcal{M}:s_d(m)\to\mathrm{Loc}_a(x,y),
\label{eq}
\end{equation}
where $s_d(m)$ represents the processed signal received at location $\alpha$, $\mathcal{T}$ represents an optional transformation operation applied to the signal $s(n)$ (such as STFT), $\mathcal{D}$ represents downsampling operation which is designed to optimize computational efficiency, the operations $\mathcal{T}$ and $\mathcal{D}$ can be interchanged, and $\operatorname{Loc}_{\alpha}(x,y)$ represents the coordinate of location $\alpha$. %If no transformation is applied, $T$ is considered the identity mapping.
To achieve accurate positioning, the objective is to minimize the error between the predicted location and the true location:
\begin{equation}
% \min | | \mathcal{M}_{\rm FM} ( s_ {\beta} ( n ) ) - \operatorname{Loc}_ { \beta } ( x , y ) | |,
\min\parallel\mathcal{M}(s_{\beta}(n))-\mathrm{Loc}_{\beta}(x,y)\parallel.
\label{eq3}
\end{equation}
In this paper, we employ deep learning methods to establish the mapping relationship.

\section{Methodology}
\label{sec3}

\subsection{Overall Framework}

\begin{figure*}[tp]
    \centering
    \includegraphics[width=0.8 \textwidth]{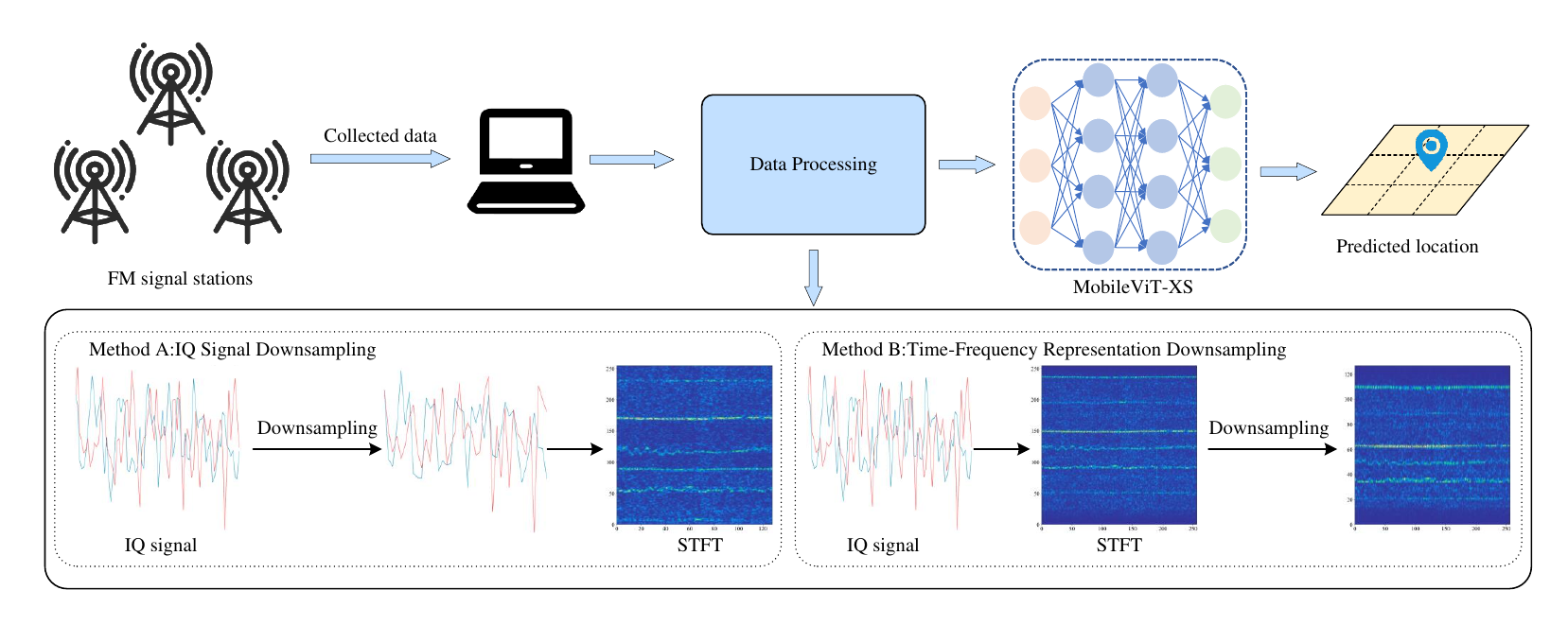}
    \caption{The overall framework of the proposed positioning method DS-Pnet.}
    \label{fig:DS-Pnet}
\end{figure*}

The overall framework of the proposed DS-Pnet is illustrated in Fig. \ref{fig:DS-Pnet}. This framework employs two downsampling methods, IQ downsampling and time-frequency downsampling, to preprocess FM signals. These methods achieve effective feature extraction while optimizing computational complexity, enabling accurate and efficient positioning by learning from downsampled data. Specifically, in Method A, the original IQ signals are downsampled to reduce data size, followed by STFT to generate time-frequency representations. In Method B, the IQ signals are first transformed into time-frequency representations using STFT, and the resulting representations are subsequently downsampled to compress redundant information while preserving essential frequency-domain features.

%DS-Pnet consists of two phases: the offline training phase and the online testing phase. In the offline training phase, FM signals are first received from multiple FM signal transmission towers within the target area to construct the dataset. Then, key features are extracted using two different downsampling strategies. In Method A, the original IQ signals are downsampled to reduce the data size, followed by Short-Time Fourier Transform (STFT) to obtain the time-frequency representation. In Method B, the original IQ signal undergoes STFT to generate the time-frequency representation, which is then downsampled to compress redundant information while retaining key frequency domain features. The downsampled time-frequency representations are fed into the MobileViT-XS model, where deep learning training produces a high-performance classification model.

DS-Pnet operates in two main phases: the offline training phase and the online testing phase. During the offline training phase, FM signals are collected from multiple transmission towers within the target area to construct the dataset. Key features are extracted using two distinct downsampling strategies. Downsampled time-frequency representations are input into the MobileViT-XS model, where deep learning techniques are applied to train a high-performance classification model. In the online testing phase,  the newly received FM signals undergo preprocessing and are analyzed with the trained classification model. A weighted positioning method is then employed to enhance flexibility, overcoming the limitation of selecting positions strictly within predefined coordinates, thereby enabling more versatile positioning capabilities.

%In summary, DS-Pnet combines downsampling methods and deep learning techniques to significantly reduce computational load while retaining effective features, thus achieving accurate and fast FM signal positioning. This downsampling-based positioning approach not only has potential applications in resource-constrained environments but also demonstrates a good balance between positioning accuracy and computational efficiency.

In summary, DS-Pnet integrates advanced downsampling methods with deep learning techniques to effectively reduce computational load while preserving essential features, achieving accurate and rapid FM signal positioning. This approach not only demonstrates strong potential in resource-constrained environments but also achieves a favorable balance between positioning accuracy and computational efficiency.

\subsection{Downsampling Methods}
In FM-Pnet, the time-frequency representation of the signal is generated using the STFT and subsequently serves as the input to a deep neural network for location estimation. Given that the computational complexity of neural networks grows with the size of the input data, reducing the dimensionality of the input signal becomes an effective strategy for lowering the network’s computational burden. However, as the downsampling factor increases, while computational complexity decreases, the positioning performance of the system tends to degrade. Consequently, it is critical to strike a balance—minimizing the network's computational complexity without causing significant deterioration in positioning accuracy. In DS-Pnet, we apply downsampling operations to both the original IQ signal and its time-frequency representation to achieve this goal.

\subsubsection{IQ Signal Downsampling}
The downsampling factor is set to $D$, meaning that one point is selected for every $D$ sampling points, resulting in a new signal. Two downsampling methods are used for downsampling the IQ signal: direct downsampling (DD) and averaging downsampling (AD).

%\paragraph{DD} 
DD is achieved by retaining the first sample point in every $D$ sample points, which can be expressed as:
\begin{equation}
y_{_{\mathrm{DD}}}(n)=x(nD),n=0,1,\ldots,L / D -1,
\label{eq4}
\end{equation}
where $x(n)$ is the original IQ signal, $y_{_{\mathrm{DD}}}(n)$ is the direct downsampled signal, and $L$ is the length of the original IQ signal.

%\paragraph{AD}
In AD, the average value of every $D$ sample points is taken as the output, which can be expressed as:
\begin{equation}
y_{\mathrm{AD}}(n)=\frac{1}{D}\sum_{m=0}^{D-1}x(nD+m),n=0,1,\ldots,L / D-1,
\label{eq5}
\end{equation}
where $y_{_{\mathrm{AD}}}(n)$ is the average downsampled signal. 

In both methods, the original IQ signal is first downsampled, after which the STFT is applied to the downsampled signal to perform a time-frequency transformation. This process produces a time-frequency representation with the time dimension reduced by a factor of $D$.

%With both methods, after downsampling the original IQ signal, the STFT is applied to the downsampled signal for time-frequency transformation, resulting in a time-frequency representation that is reduced by a factor of $D$ in the time dimension.

\begin{figure*}[tp]
    \centering
    \includegraphics[width=0.85 \textwidth]{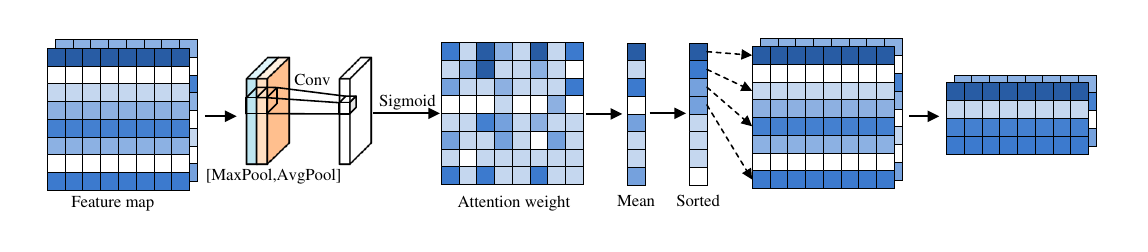}
    \caption{The attention mechanism-based downsampling framework.}
    \label{fig:sam}
\end{figure*}

%\subsubsection{Time-Frequency Representation Downsampling}
\subsubsection{Time-Frequency Direct Downsampling}

The application of STFT produces a time-frequency representation in the form of a complex matrix with dimensions $p \times q$. To fully capture the information contained within, the real and imaginary parts of the matrix are extracted separately. This results in the final time-frequency representation $M_{s}$, ${M_s} \in {\Re ^{2 \times p \times q}}$. For time-frequency representation $M_{s}$, we apply two direct downsampling methods: max pooling downsampling (MPD) and average pooling downsampling (APD).

%\paragraph{MPD}
MPD applies a max pooling layer to iteratively select the maximum value over the pooling window on the input time-frequency representation $M_{s}$, expressed as:
\begin{equation}
S_{_{\operatorname{Max}}} = \operatorname{MaxPool2d} (M_{_s}),
\label{eq6}
\end{equation}
where $S_{_{Max}}$ is the max pooling downsampled time-frequency representation. When downsampling along the time dimension, the kernel size is $(1,D)$ with a stride of $(1,D)$. When downsampling along the frequency dimension, the kernel size of the pooling layer is $(D,1)$ with a stride of $(D,1)$.

%\paragraph{APD}
APD applies an average pooling layer to iteratively compute the mean value over the pooling window on the input time-frequency representation $M_{s}$, which can be expressed as:
\begin{equation}
S_{_{\operatorname{Avg}}}=\operatorname{AvgPool2d}(M_s),
\label{eq7}
\end{equation}
where $S_{_{Avg}}$ is the average pool downsampled time-frequency representation. The operation, where downsampling is applied along the time dimension with a kernel size of $(1,D)$ and a stride of $(1,D)$ , is referred to as APD-T, while downsampling along the frequency dimension is referred to as APD-F.

% and attention mechanism-based downsampling (AMD)
%\paragraph{AMD}
\subsubsection{Time-Frequency Attention Downsampling}

During the direct downsampling process, while max pooling and average pooling can effectively reduce computational complexity, they may result in information loss due to their fixed-window nature. To address this issue, we propose a time-frequency attention downsampling method that incorporates a spatial attention mechanism (SAM) \cite{27}, called attention mechanism-based downsampling (AMD). Considering that FM signals within the sampled bandwidth typically consist of a small number of narrowband FM signals and a large amount of noise subbands, these narrowband signals exhibit sparsity in the time-frequency representation. By introducing the attention mechanism, the network can focus more on the regions of the time-frequency representation with higher signal strength. In the attention weight matrix, the weights in signal regions are closer to 1, reflecting the importance of these regions for the localization task. Based on this characteristic, we process the attention weight matrix and selectively retain the important feature rows. This enables the model to more easily learn key features during training, achieving downsampling while minimizing the interference of noise.

The proposed downsampling method combined with the spatial attention mechanism is illustrated in Fig. \ref{fig:sam}. The spatial attention mechanism first applies average pooling and max pooling along the channel dimension to the time-frequency representation, extracting the maximum and average values. Then, the average and maximum value matrices are concatenated along the channel dimension. The combined matrix is then processed through a 2D convolution layer, which integrates information across channels to produce a feature map with a single channel. Finally, the attention weight matrix is generated through the Sigmoid function. The process can be expressed as: 
\begin{equation}
Q_1=\sigma(\text{Conv2D}([\text{MaxPool}(\text{M}_s);\text{AvgPool}(\text{M}_s)])),
\label{eq8}
\end{equation}
where $\sigma( \cdot )$ represents the Sigmoid function, ${Q_1} \in {\Re ^{p \times q}}$  represents the attention weight map, $\operatorname{Conv2D}( \cdot )$  refers to a 2D convolutional layer with 2 input channels and 1 output channel. The layer uses a kernel size of $7 \times 7$, a stride of 1, and a padding of 3.

To evaluate the importance of each row's information for the task, we take the row-wise average of the attention weight matrix ${Q_1}$ and sort the rows based on the average values. According to the set downsampling factor, we retain the row indices corresponding to the frequencies in the time-frequency representation. Then, based on these indices, we select the corresponding frequency rows from the time-frequency representation and arrange these rows in their original order. The process can be expressed as:
\begin{equation}
\overline{Q}_{1}(i)=\frac{1}{N}\sum_{j=1}^{N}Q_{1}(i,j),
\label{eq9}
\end{equation}
\begin{equation}
S=\operatorname{top-k}(\overline{Q}_1,k=\frac{N}{D}),
\label{eq10}
\end{equation}
where $\overline{Q}_{1}(i)$ represents the average weight of the $i$-th row, $N$ is the number of columns in the attention weight matrix ${Q_1}$, $S$  represents the frequency rows selected based on the average weight values in descending order, $k$ is the number of retained frequency rows, and $D$ is the downsampling factor.

During the training process, we incorporate the spatial attention-based downsampling method into the neural network for joint training. Considering that the frequency rows retained for each sample during training may differ when using the attention mechanism for downsampling, we establish a validation set after training to track the most frequently discarded frequency rows. Based on the downsampling factor, we select and save the indices of the discarded frequency rows, and during training, this index file is used to construct a downsampled dataset for re-training the network. During the re-training process, the attention mechanism is not needed, and the downsampling is performed based solely on the indices of the discarded frequency rows from the index file. At this stage, the attention mechanism-based downsampling does not add extra computational complexity to the network.

\begin{figure*}[tp]
\centering
\subfigure[]{\label{fig:indoor:ab}
\includegraphics[width=0.2\linewidth, height=3.5cm]{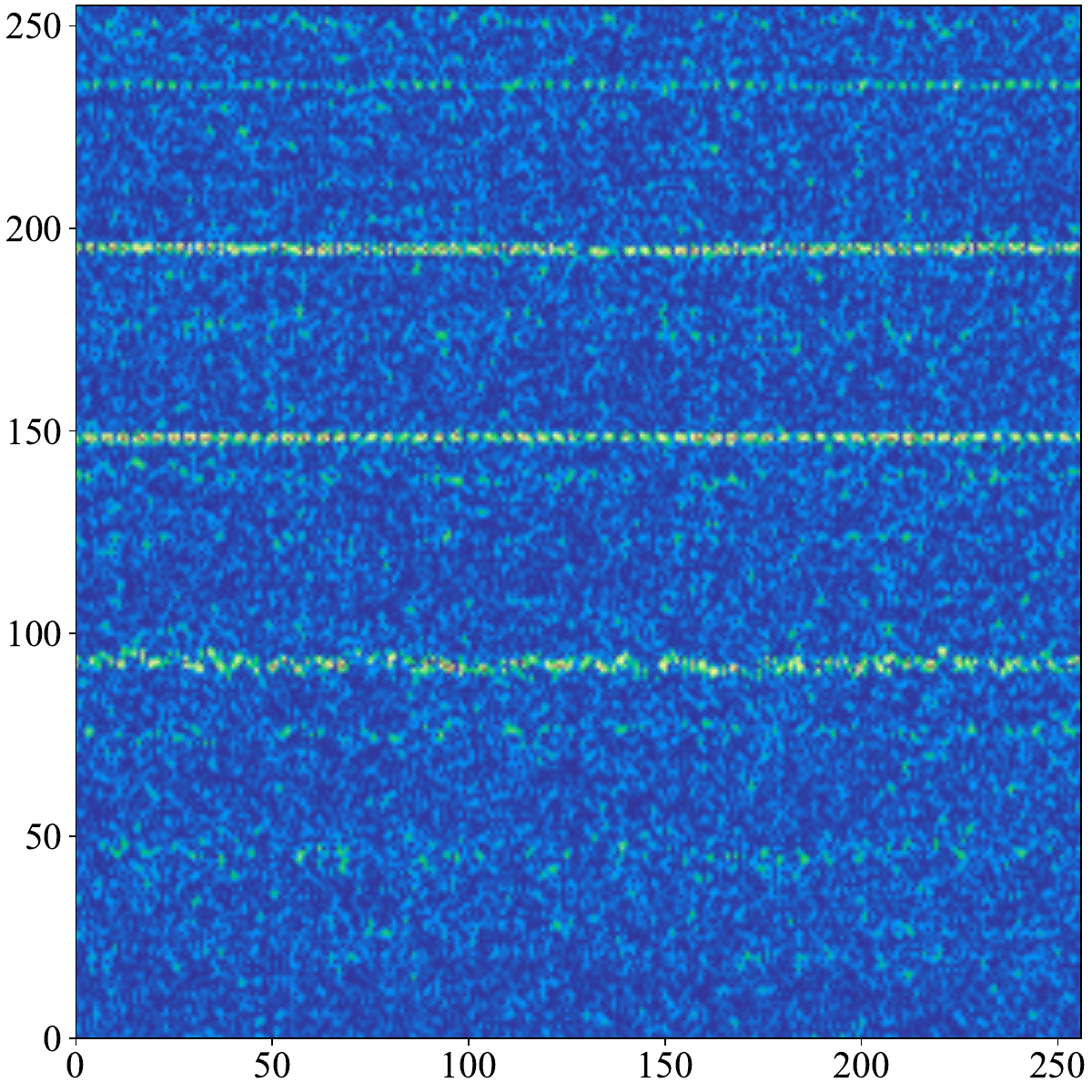}}
\hspace{0.02\linewidth}
\subfigure[]{\label{fig:indoor:bc}
\includegraphics[width=0.2\linewidth, height=3.5cm]{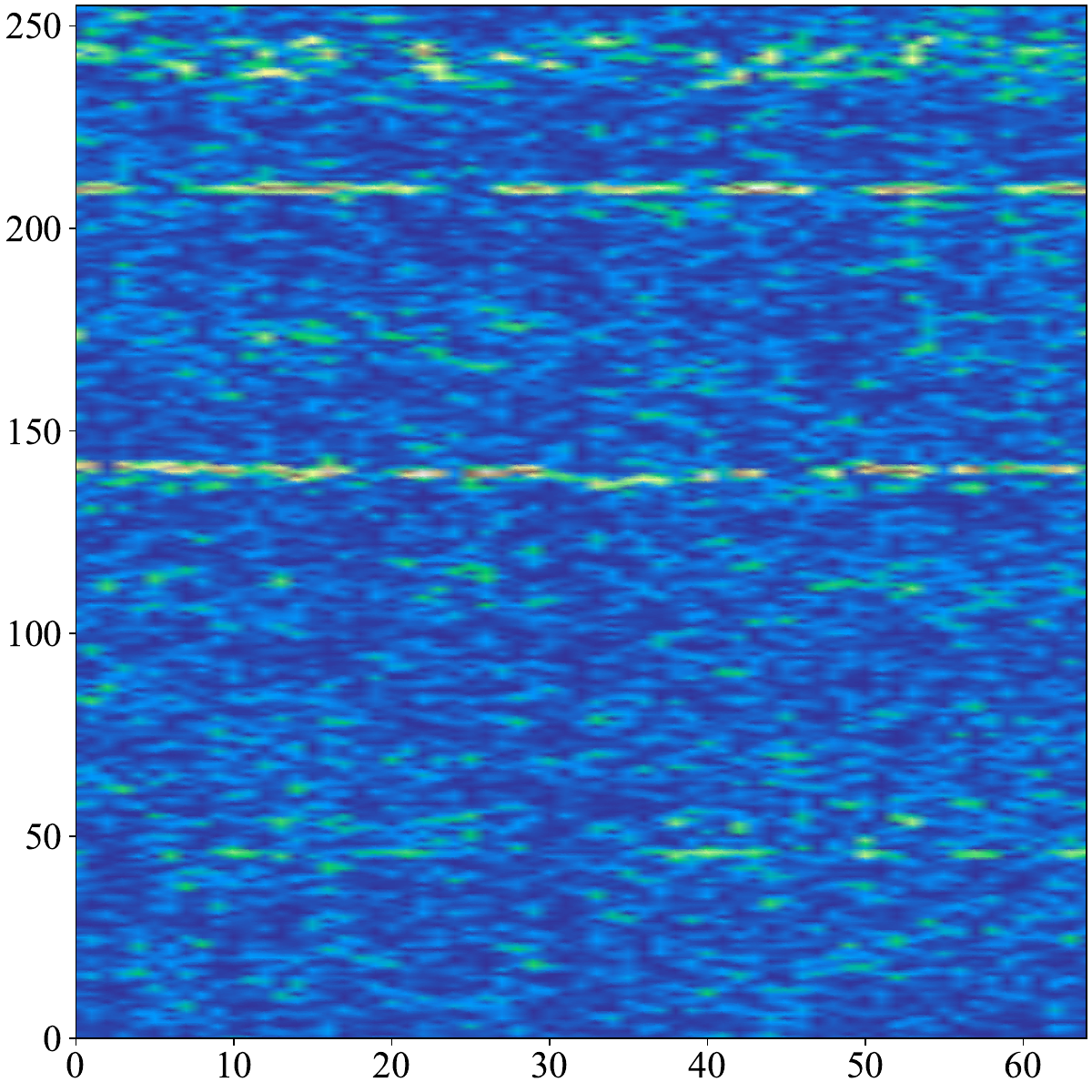}}
\hspace{0.02\linewidth}
\subfigure[]{\label{fig:indoor:cd}
\includegraphics[width=0.2\linewidth, height=3.5cm]{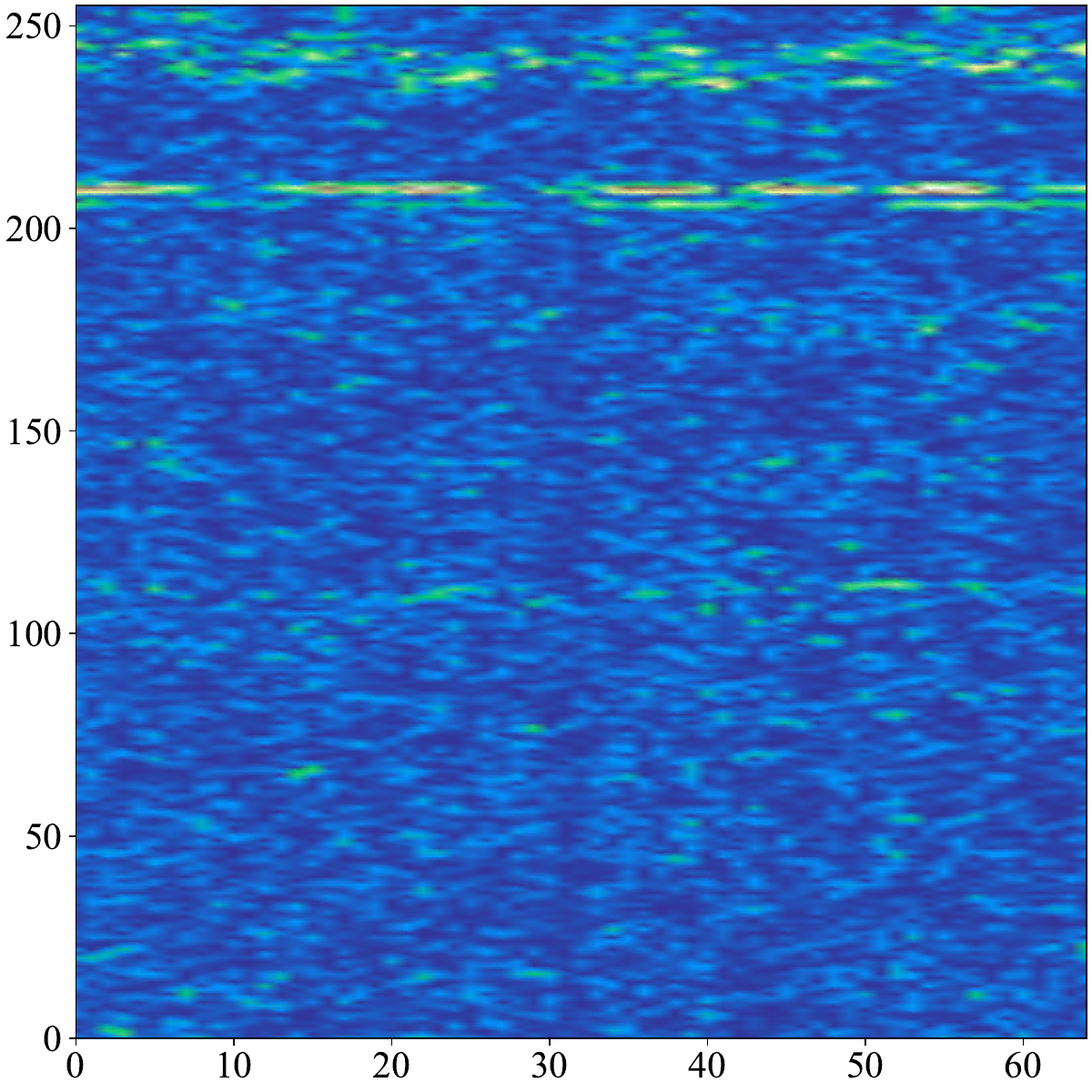}}
\hspace{0.02\linewidth}
\subfigure[]{\label{fig:indoor:de}
\includegraphics[width=0.2\linewidth, height=3.5cm]{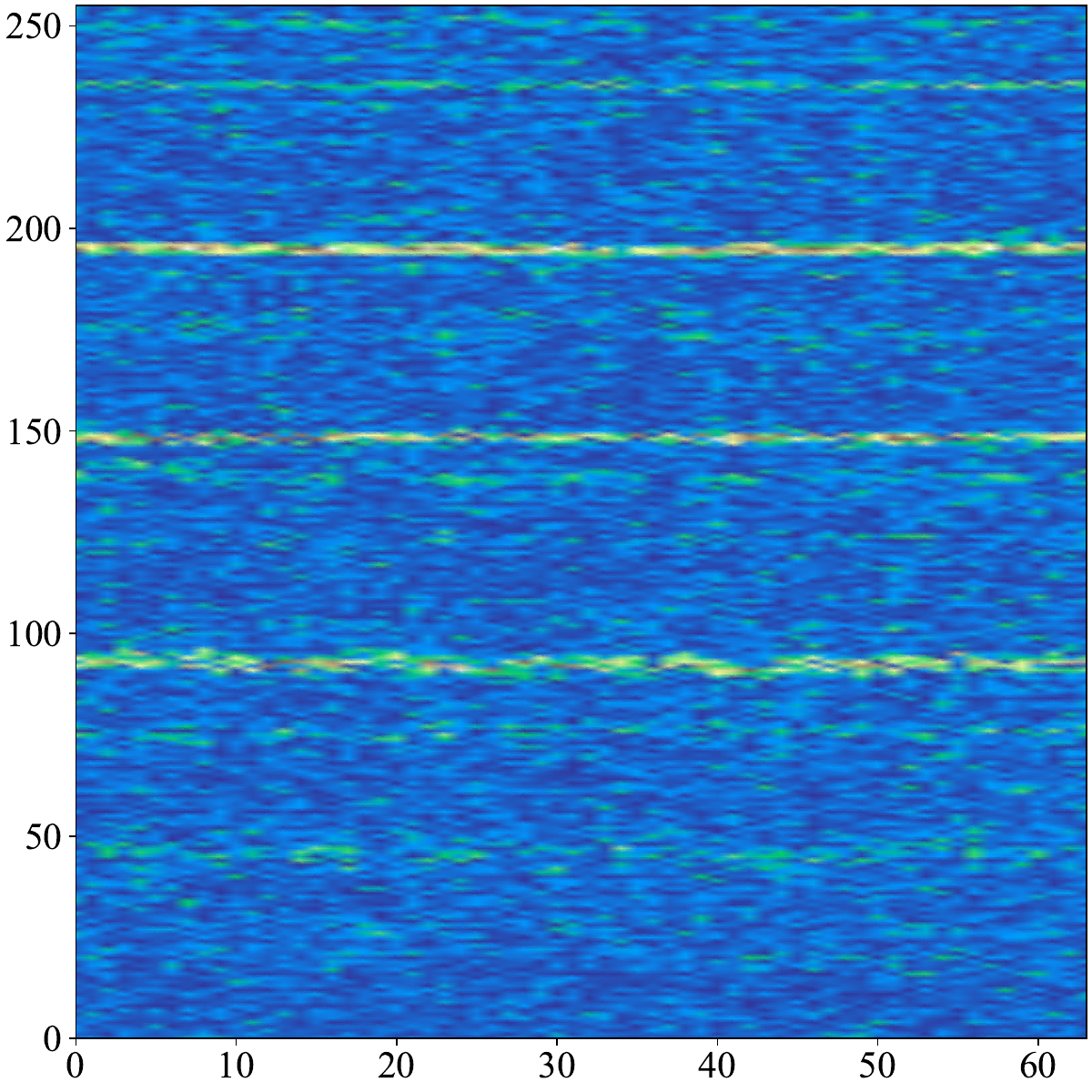}}
\hspace{0.02\linewidth}
\subfigure[]{\label{fig:indoor:ef}
\vspace{2mm}
\includegraphics[width=0.2\linewidth, height=3.5cm]{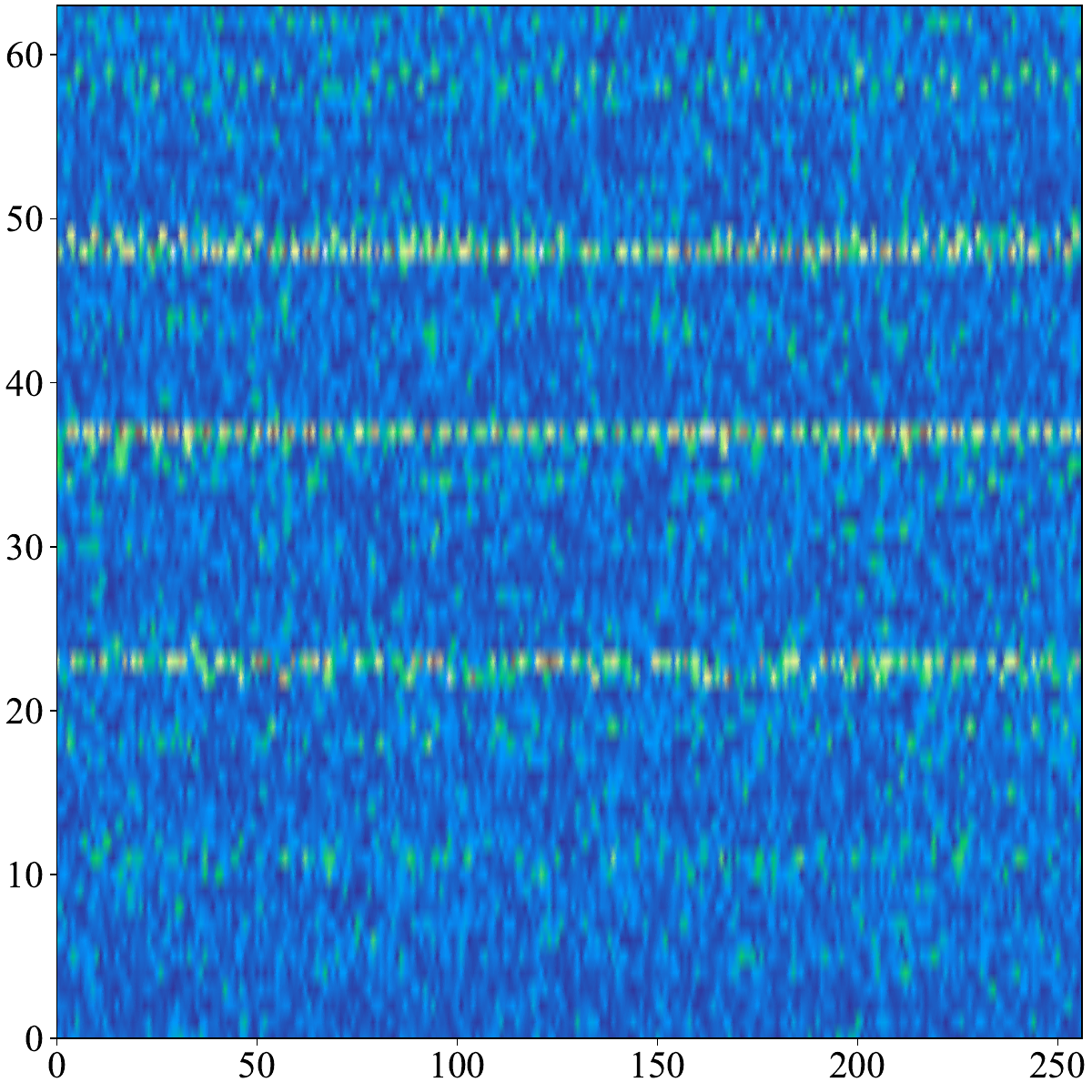}}
\hspace{0.02\linewidth}
\subfigure[]{\label{fig:indoor:fg}
\includegraphics[width=0.2\linewidth, height=3.5cm]{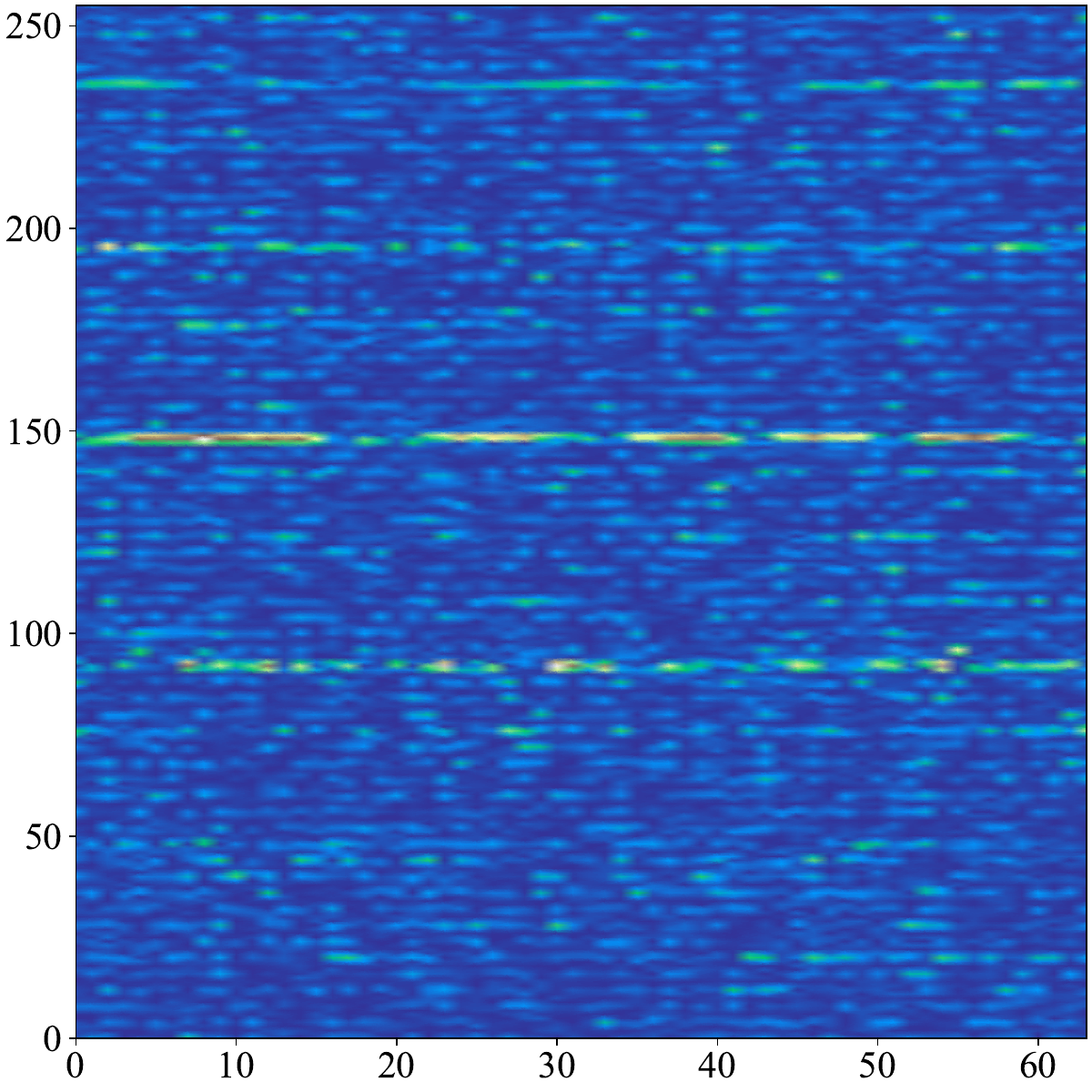}}
\hspace{0.02\linewidth}
\subfigure[]{\label{fig:indoor:gh}
\includegraphics[width=0.2\linewidth, height=3.5cm]{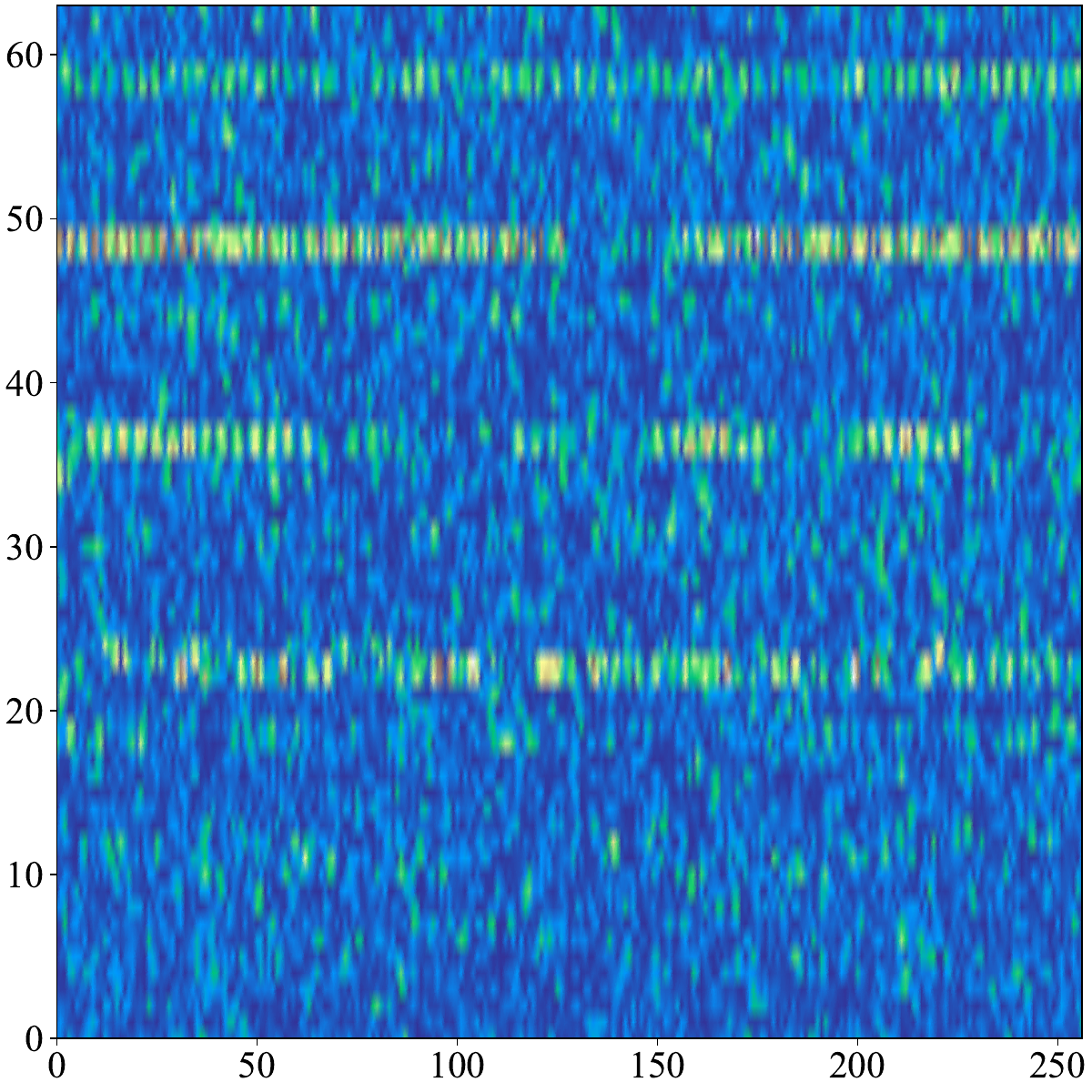}}
\hspace{0.02\linewidth}
\subfigure[]{\label{fig:indoor:ha}
\includegraphics[width=0.2\linewidth, height=3.5cm]{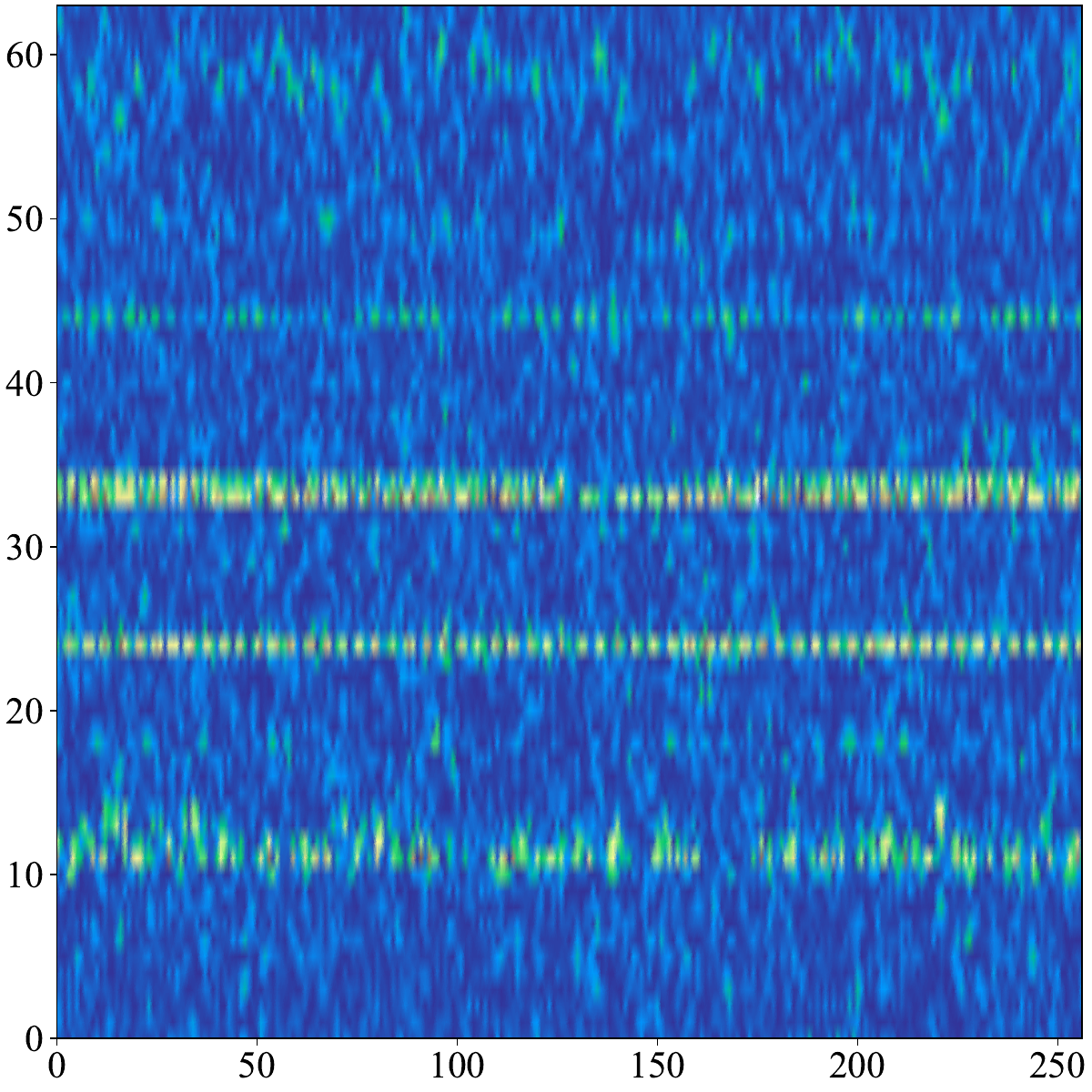}}
% \subfigure[]{\label{fig:indoor:ca}
% \includegraphics[width=0.32\linewidth]{figs/1012.eps}}   
\centering
\caption{The time-frequency representation with 4x downsampling factor. (a) Original, (b) AD, (c) DD, (d) MPD-T, (e) MPD-F, (f) APD-T, (g) APD-F, and (h) AMD.}
\label{fig:STFT}
\end{figure*}

\subsubsection{Illustration of Downsampling}
The time-frequency representations under a 4x downsampling factor using different downsampling methods are shown in Fig. \ref{fig:STFT}. In the figure, the horizontal axis represents time, and the vertical axis represents frequency, with various colors indicating different values in the FM signal sample sequence. Brighter areas correspond to regions of higher signal strength, typically associated with high-energy areas, while darker areas indicate low energy, usually representing background noise. As shown, AMD retains the most features from the original time-frequency representation compared to other methods.

The aforementioned downsampling methods only change the size of the input data, so there is no need to modify the network structure or the output length of the network, only the input length needs to be adjusted. However, we need to find a balance that reduces computational complexity without significantly degrading the system's localization performance.

% \begin{figure*}[tp]
%     \centering
%     \includegraphics[width=0.8 \textwidth]{new-fig/MobileVIT.pdf}
%     \caption{The network structure of MobileViT-XS.}
%     \label{fig:VIT}
% \end{figure*}
\begin{figure*}[tp]
    \centering
    \includegraphics[width=0.8 \textwidth]{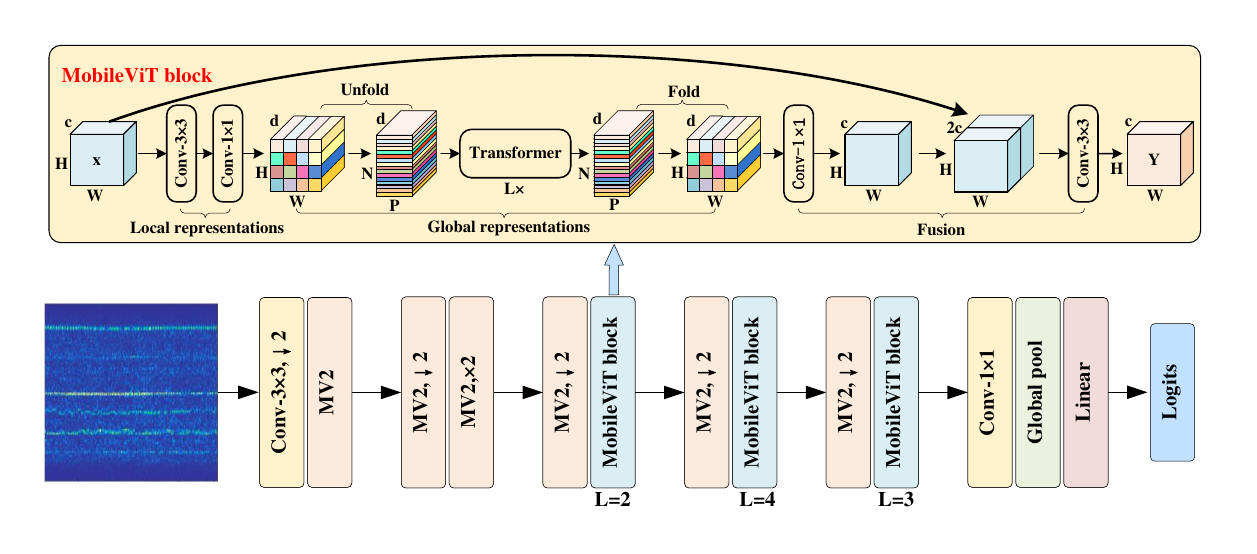}
    \caption{The network structure of MobileViT-XS.}
    \label{fig:VIT}
\end{figure*}
\subsection{Adopted Network Structure}
To strike a balance between computational efficiency and model localization performance, we adopt a hybrid architecture, MobileViT-XS \cite{28}. MobileViT-XS combines the advantages of lightweight convolutional neural networks (CNNs) and Transformers to enhance both model performance and efficiency. The architecture consists of standard convolutions, MV2 (Inverted Residual Block in MobileNet\_v2) \cite{29}, MobileViT block, global pooling, and fully connected layers. Among these, the MV2 block serves as the backbone convolution module, extracting local features through expansion, compression, and residual connections, thereby reducing computational complexity. However, traditional CNNs primarily focus on learning local features, which results in the limited effective receptive field of deep networks, preventing them from covering the entire input image and constraining their ability to model global features. To address this limitation, MobileViT-XS introduces the MobileViT block for effective global feature extraction.The MobileViT block consists of three sub-modules: the local representation module, the global representation module, and the fusion module. The local representation module extracts local features through convolutional layers, while the global representation module extracts global features through unfolding, local processing, and folding operations. In particular, the local processing part utilizes the Transformer mechanism, which can capture long-range dependencies between feature blocks, thus forming feature representations that include global context. The fusion module combines the globally remapped features with the initial convolution features, allowing the output features to contain both local details and global context, thereby improving the model’s ability to recognize complex image patterns.

The MobileViT-XS network structure used in this paper is shown in Fig. \ref{fig:VIT}. Convolutional layers and the convolution operations within the MobileViT blocks are used to extract local features and achieve information fusion along the channel dimension. The MobileViT block simultaneously learns both local and global representations and is placed in the deeper layers of the network, alternating with  MobileNet\_v2 blocks to form the core structure of the MobileViT-XS network. In the architecture, ↓2 represents downsampling, and L denotes the number of Transformer layers in each MobileViT block. Stacking multiple layers helps the model extract more powerful global information without significantly increasing the number of parameters. Finally, the classification result is obtained through a  $1 \times 1$ convolution, a global average pooling layer, and a fully connected layer.

%表1
% \begin{table}
% \caption{Basic Structure of the Network in FM-Pnet}
% \centering
% \renewcommand\arraystretch{1.0}
% \setlength{\tabcolsep}{10mm}{
% \begin{tabular}{|c|c|}
% \hline
% Layer & Detail \\ \hline
% conv1 & $3\times3,64, \text{stride}$ \\ \hline
% conv2 & 
% $\begin{bmatrix}
% 1\times1,64 \\
% 3\times3,64\text{ ($C = 32$)} \\
% 1\times1,128
% \end{bmatrix} \times 3$ \\ \hline
% conv3 & 
% $\begin{bmatrix}
% 1\times1,128 \\
% 3\times3,128\text{ ($C = 32$)} \\
% 1\times1,512
% \end{bmatrix} \times 4$ \\ \hline
% conv4 & 
% $\begin{bmatrix}
% 1\times1,256 \\
% 3\times3,256\text{ ($C = 32$)} \\
% 1\times1,512
% \end{bmatrix} \times 5$ \\ \hline
% conv5 & 
% $\begin{bmatrix}
% 1\times1,512 \\
% 3\times3,512\text{ ($C = 32$)} \\
% 1\times1,1024
% \end{bmatrix} \times 6$ \\ \hline
% conv6 & 
% $\begin{bmatrix}
% 1\times1,1024 \\
% 3\times3,1024\text{ ($C = 32$)} \\
% 1\times1,2048
% \end{bmatrix} \times 3$ \\ \hline
% pooling& Adaptive Average Pool \\ \hline
% FC& $G$-d fc, softmax \\ \hline
% \end{tabular}}
% \label{table1}
% \end{table}
%$\rho_i (\mathcal{S}|{f_\theta })$
\subsection{The Training Algorithm}
In model training, the loss function plays a crucial role as it measures the difference between the model's predictions and the actual labels, guiding the model towards convergence during the training process. This paper uses the cross-entropy loss function, defined as follows:
\begin{equation}
\ell=-\sum_{i=1}^{C}y_i\log(p_i),
\label{eq11}
\end{equation}
where $C$ is the total number of classes, ${y_i}$ represents the true label of the sample, and $p_i $ represents the probability predicted by the model that the sample belongs to class $i$. For a batch of data, the loss function is computed independently for each sample, and then the average value is taken to obtain the overall loss for the batch. The AdamW \cite{30} optimizer is used in model training. Compared to the classic Adam optimizer, AdamW directly applies weight decay, which effectively controls the update of model parameters, helping to alleviate overfitting and improve training stability. AdamW performs exceptionally well in the training of deep learning models, making it particularly suitable for complex tasks that require long-term optimization. Throughout the training process, the network parameters are continuously adjusted using the cross-entropy loss function and the AdamW optimizer to optimize the model's performance. After the model training is completed, to improve the accuracy of location estimation, we further optimize the prediction results during the testing phase through Bayesian re-evaluation, rather than solely relying on the direct output of the trained model \cite{31}. This process involves preprocessing the received FM signal, then feeding the processed data into the model to obtain the predicted location of the target, achieving high-precision estimation of the target's position.

\section{Experiment}
\label{sec4}
\subsection{Experimental Setup}
\subsubsection{Dataset and Parameter Settings}

The dataset utilized in this paper is sourced from \cite{22} , employing a 4 MHz bandwidth signal dataset for analysis. To account for the impact of downsampling on signal length and to evaluate the effectiveness of the downsampling approach, the signal sample length is set to 16,384. With this sample length, the total number of samples is relatively small. Consequently, each dataset collected from a given point comprises 2,000 training samples and 1,000 test samples.
Furthermore, the configuration of short-time Fourier transform (STFT) parameters significantly influences the performance of DS-Pnet. To ensure consistency and comparability across experiments, the STFT parameters are fixed as follows: the FFT size is set to 256, the overlap ratio is maintained at 0.75, and the Hanning window is utilized as the window function.

\subsubsection{Operating Environment}
All experiments were conducted using an NVIDIA GeForce RTX 4060 GPU and the PyTorch framework. During training, the AdamW optimization algorithm was used for parameter updates, with a batch size of 8 and an initial learning rate of 0.001. The learning rate was halved every two epochs, and a total of 15 training epochs were completed.

% \subsubsection{Baselines}
% We conduct a comparative analysis of our proposed method against existing techniques, namely deterministic KWNN \cite{6199864}, the probability-based histogram method (Histogram) \cite{1424614}, and the Combined method \cite{moghtadaiee2014indoor}, which combines the KWNN and Histogram approaches. The Combined method entails computing the arithmetic average of the positioning prediction outcomes from both the KWNN and Histogram methods to derive its ultimate positioning result. For the KWNN method, we iterate over all possible values of $K$ and select the one that yields optimal positioning performance. In the Histogram method, we determine the number of bins for calculating the histogram as the square root of the number of training samples. These three existing methods necessitate the use of RSS from FM signals as both training and testing samples for positioning. The number of channels for RSS calculation varies with the collected bandwidth \cite{moghtadaiee2014indoor}: 1 channel for the 320kHz bandwidth, 5 channels for the 4MHz bandwidth, and 22 channels for the 20MHz bandwidth. Following the calculation, we obtain RSS vectors with dimensions of 1, 5, and 22, respectively. In the experiments, random guessing is deemed as unsuccessful positioning, and the error associated with random guessing varies across different-sized areas.

\subsubsection{Performance Metrics}
To evaluate the effectiveness of the proposed method, three performance metrics are considered: mean distance error (MDE), standard deviation (STD), and cumulative distribution function (CDF).

%\paragraph{Mean Distance Error (MDE)}
% MDE is a method used to measure the accuracy of predictions made by statistical models. It is calculated by taking the average of the squared differences between the predicted values and the actual observations. 
MDE  is an important metric for evaluating the accuracy of a localization system, as it effectively quantifies the difference between the predicted and true positions. The formula for MDE is expressed as:
\begin{equation}
\begin{aligned}
\mathrm{MDE}=\frac{1}{M}\sum_{i=1}^{M}\|\mathbf{x}_i-\mathbf{\hat{x}}_i\|,
\end{aligned}
\label{eq12}
\end{equation}
where  $M$ represents the number of test samples, $\mathbf{x}_i$ The true position vector of the  $i$-th sample, $\mathbf{\hat{x}}_i$ The predicted position vector of the 
 $i$-th sample, and  $\left\|\cdot\right\| $ represents the Euclidean distance between two position vectors.

%\paragraph{Standard Deviation (STD)}
STD is a statistical metric that measures the dispersion of a dataset and can quantify the variation in test distance errors using standard deviation:
\begin{equation}
\begin{aligned}
\operatorname{STD}  = \sqrt {\frac{1}{M}\sum\limits_{i = 1}^M {({Y_i} - \operatorname{MDE}} {)^2}},
\end{aligned}
\label{eq13}
\end{equation}
where $Y_i$ %$D_i=\sqrt {{{(x_i - {\hat{x}_i})}^2} + {{(y_i - {\hat{y}_i})}^2}}$
denotes the distance error for the $i$-th sample.  
A smaller STD indicates that the localization errors are more concentrated around the mean, and the system's localization results are more stable. A larger STD suggests that the localization results exhibit higher variability.

%\paragraph{Cumulative Distribution Function (CDF)}
% \textbf{Cumulative Distribution Function (CDF)}. 
To comprehensively describe the probability distribution of the overall error, CDF is introduced, which is expressed as:
\begin{equation}
\begin{aligned}
\operatorname{CDF}\left(\mid error\mid\right)=P(X<\mid error\mid),
\end{aligned}
\label{eq14}
\end{equation}
where the CDF represents the sum of the probabilities of all errors less than or equal to a certain value, 
$\mid error\mid$ denotes the error between the estimated and true values, and $X$ represents the set of all errors between the estimated and true values. By analyzing the CDF curve, one can intuitively understand the probability distribution of the localization system within different error ranges, thus evaluating the system's performance in meeting various accuracy requirements.

\subsection{The Impact of Different Networks}
% \begin{figure}
% \centering
% \subfigure[]{
% \label{fig:different_overlap:a}
% \begin{minipage}[b]{0.22\textwidth}  % 0.4
% \includegraphics[width=1\textwidth]{figs/new_overlap_in.eps} 
% \end{minipage}
% }
% \subfigure[]{
% \label{fig:different_overlap:b}
% \begin{minipage}[b]{0.22\textwidth}  % 0.4
% \includegraphics[width=1\textwidth]{figs/new_overlap_out.eps} 
% \end{minipage}}
% \caption{The influence of different overlap ratios on the positioning performance of FM-Pnet. (a) Indoor scenario, (b) outdoor scenario.}
% \label{fig:different_overlap}
% \end{figure}

We first compare the performance of the lightweight neural network MobileViT-XS with the improved ResNeXt model used in FM-Pnet. As shown in Table. \ref{table1}, in indoor scenarios, the MDE 
of ResNeXt is 0.0531, while the MDE of MobileViT-XS is 0.0503, indicating similar localization accuracy. However, MobileViT-XS exhibits a lower STD value of 0.3465 compared to ResNeXt's STD value of 0.5249, suggesting that MobileViT-XS has a more concentrated error distribution and demonstrates superior localization stability in indoor environments. In outdoor scenarios, MobileViT-XS achieves an MDE of 0.7787 and an STD of 2.1418, whereas ResNeXt shows an MDE of 0.9240 and an STD of 2.7584. This indicates that MobileViT-XS offers improvements in both localization accuracy and stability in outdoor environments.

\begin{figure}[t]
    \centering
    \includegraphics[width=0.4\textwidth]{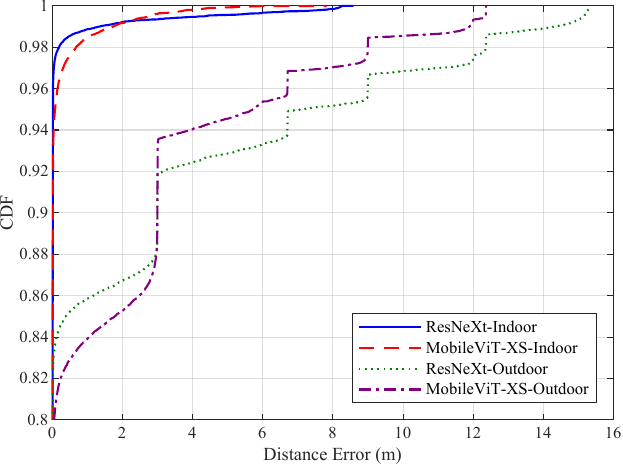}
    \caption{The impact of different networks on positioning performance.}
    \label{fig:net}
\end{figure}

The CDF results for both indoor and outdoor scenarios are presented in Fig.  \ref{fig:net}. The CDF curve of MobileViT-XS starts at a similar value to that of ResNeXt, but MobileViT-XS exhibits a faster convergence rate. In conclusion, MobileViT-XS demonstrates superior localization performance in both indoor and outdoor environments compared to ResNeXt. With its more lightweight architecture and faster computational efficiency, MobileViT-XS presents a more efficient positioning model.

\begin{table}[t]
\centering
\caption{Test Results of Different Networks}
\renewcommand\arraystretch{1.0}
\begin{tabular}{|cl|cc|cc|}
\hline
\multicolumn{2}{|c|}{\multirow{2}{*}{Model}} & \multicolumn{2}{c|}{Indoor}          & \multicolumn{2}{c|}{Outdoor}         \\ \cline{3-6} 
\multicolumn{2}{|c|}{}                       & \multicolumn{1}{c|}{MDE}    & STD    & \multicolumn{1}{c|}{MDE}    & STD    \\ \hline
\multicolumn{2}{|c|}{ResNeXt}                & \multicolumn{1}{c|}{0.0531} & 0.5249 & \multicolumn{1}{c|}{0.9240} & 2.7584 \\ \hline
\multicolumn{2}{|l|}{MobileViT-XS}           & \multicolumn{1}{c|}{0.0503} & 0.3465 & \multicolumn{1}{c|}{0.7787} & 2.1418 \\ \hline
\end{tabular}
\label{table1}
\end{table}

\subsection{Performance of Time Dimension Downsampling}
% \begin{figure}
% \centering
% \subfigure[]{
% \label{fig:different_window:a}
% \begin{minipage}[b]{0.22\textwidth}  % 0.4
% \includegraphics[width=1\textwidth]{figs/new_windows_in.eps} 
% \end{minipage}
% }
% \subfigure[]{
% \label{fig:different_window:b}
% \begin{minipage}[b]{0.22\textwidth}  % 0.4
% \includegraphics[width=1\textwidth]{figs/new_windows_out.eps} 
% \end{minipage}}
% \caption{The influence of different window functions on the positioning performance of FM-Pnet. (a) Indoor scenario, (b) outdoor scenario.}
% \label{fig:different_window}
% \end{figure}
\begin{table}[t]
\caption{Test Results of Different Downsampling Factors and Methods }
\centering
\renewcommand\arraystretch{1.0}
\setlength{\tabcolsep}{1.75mm}
\begin{tabular}{|c|c|cc|cc|}
\hline
\multirow{2}{*}{\begin{tabular}[c]{@{}c@{}}Downsampling\\  factor\end{tabular}} & \multirow{2}{*}{\begin{tabular}[c]{@{}c@{}}Downsampling \\ method\end{tabular}} & \multicolumn{2}{c|}{Indoor}          & \multicolumn{2}{c|}{Outdoor}         \\ \cline{3-6} 
                                                                                &                                                                                 & \multicolumn{1}{c|}{MDE}    & STD    & \multicolumn{1}{c|}{MDE}    & STD    \\ \hline
\multirow{4}{*}{2}                                                              & DD                                                                              & \multicolumn{1}{c|}{0.0518} & 0.3663 & \multicolumn{1}{c|}{0.7334} & 2.0564 \\ \cline{2-6} 
                                                                                & AD                                                                              & \multicolumn{1}{c|}{0.0412} & 0.3195 & \multicolumn{1}{c|}{0.7184} & 2.0525 \\ \cline{2-6} 
                                                                                & MPD                                                                             & \multicolumn{1}{c|}{0.0989} & 0.5771 & \multicolumn{1}{c|}{0.7794} & 2.2174 \\ \cline{2-6} 
                                                                                & APD                                                                             & \multicolumn{1}{c|}{0.0494} & 0.3903 & \multicolumn{1}{c|}{0.9517} & 2.6738 \\ \hline
\multirow{4}{*}{4}                                                              & DD                                                                              & \multicolumn{1}{c|}{0.0527} & 0.3891 & \multicolumn{1}{c|}{1.0611} & 2.5695 \\ \cline{2-6} 
                                                                                & AD                                                                              & \multicolumn{1}{c|}{0.1965} & 0.8566 & \multicolumn{1}{c|}{0.8116} & 2.0588 \\ \cline{2-6} 
                                                                                & MPD                                                                             & \multicolumn{1}{c|}{0.0259} & 0.2102 & \multicolumn{1}{c|}{0.7628} & 2.0970 \\ \cline{2-6} 
                                                                                & APD                                                                             & \multicolumn{1}{c|}{0.0472} & 0.3755 & \multicolumn{1}{c|}{0.9257} & 2.7310  \\ \hline
\multirow{4}{*}{8}                                                              & DD                                                                              & \multicolumn{1}{c|}{0.0304} & 0.3134 & \multicolumn{1}{c|}{1.7304} & 3.2955 \\ \cline{2-6} 
                                                                                & AD                                                                              & \multicolumn{1}{c|}{0.2137} & 0.8304 & \multicolumn{1}{c|}{1.5146} & 2.4509 \\ \cline{2-6} 
                                                                                & MPD                                                                             & \multicolumn{1}{c|}{0.0274} & 0.2405 & \multicolumn{1}{c|}{0.8143} & 2.3314 \\ \cline{2-6} 
                                                                                & APD                                                                             & \multicolumn{1}{c|}{0.0602} & 0.3715 & \multicolumn{1}{c|}{0.8488} & 2.2765 \\ \hline
\multirow{4}{*}{16}                                                             & DD                                                                              & \multicolumn{1}{c|}{0.2475} & 0.9767 & \multicolumn{1}{c|}{3.8825} & 4.1268 \\ \cline{2-6} 
                                                                                & AD                                                                              & \multicolumn{1}{c|}{0.9048} & 1.7633 & \multicolumn{1}{c|}{3.2881} & 3.0208 \\ \cline{2-6} 
                                                                                & MPD                                                                             & \multicolumn{1}{c|}{0.0256} & 0.2329 & \multicolumn{1}{c|}{0.7148} & 2.0949 \\ \cline{2-6} 
                                                                                & APD                                                                             & \multicolumn{1}{c|}{0.1689} & 0.6689 & \multicolumn{1}{c|}{1.1047} & 2.6871 \\ \hline
\end{tabular}
\label{table2}
\end{table}

% Finally, we investigate the impact of different window function choices on the positioning performance, including Hanning, Hamming, Blackman, and Triangular. As depicted in Fig. \ref{fig:different_window:a}, using the Hanning window for indoor positioning yields the best performance. With the Hanning window, over 96\% of the sample testing errors are within 0.05, showcasing a significant advantage compared to the other three windows. Additionally, the performance of Blackman and Hamming windows for positioning is similar, while the Triangular window exhibits the poorest performance. For outdoor scenario, as shown in Fig. \ref{fig:different_window:b}, using the Hanning window results in over 90\% of the samples having distance errors less than 5, surpassing the results obtained with the other three window functions. The results from both indoor and outdoor scenarios indicate that the performance of FM-Pnet is optimal when the Hanning window is chosen. We will use Hanning window in the following experiments. 
\begin{figure*}[t]
\centering
\subfigure[]{\label{fig:indoor-s-2:ab}
\includegraphics[width=0.4\linewidth]{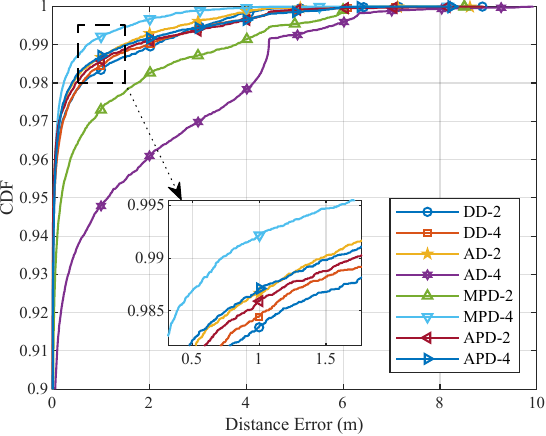}}
\hspace{0.02\linewidth}
\subfigure[]{\label{fig:indoor-s-8:b}
\includegraphics[width=0.4\linewidth]{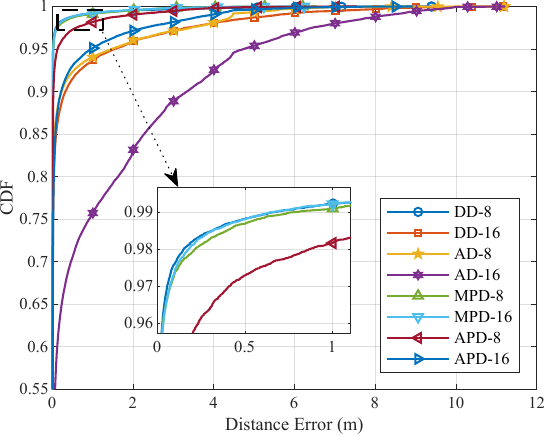}}
\vspace{2mm}
\hspace{0.02\linewidth}
\subfigure[]{\label{fig:outdoor-s-2:c}
\includegraphics[width=0.4\linewidth]{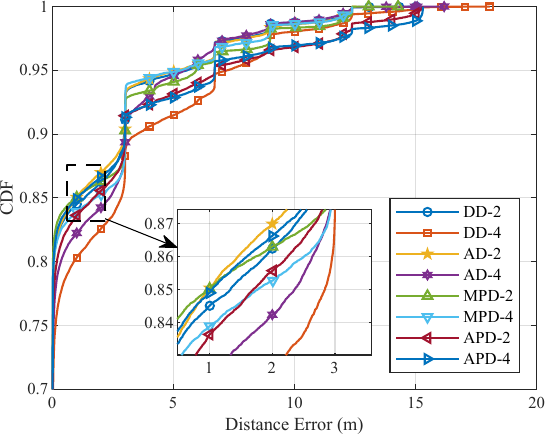}}
\hspace{0.02\linewidth}
\subfigure[]{\label{fig:outdoor-s-8:d}
\includegraphics[width=0.4\linewidth]{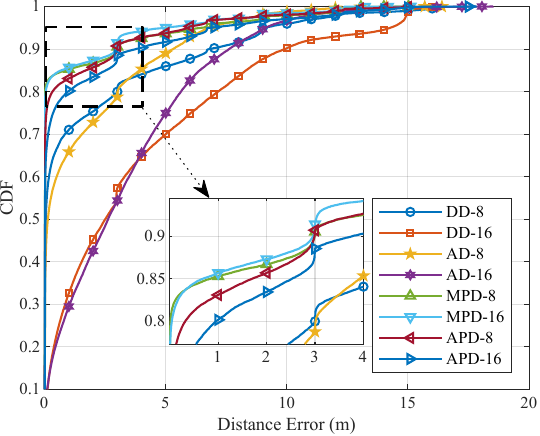}}
\hspace{0.02\linewidth}
% \subfigure[]{\label{fig:indoor:ca}
% \includegraphics[width=0.32\linewidth]{figs/1012.eps}}   
\centering
\caption{The Impact of Different Downsampling Factors and Methods on Localization Performance. (a) Indoor 2,4, (b) Indoor 8,16, (c) Outdoor 2,4, (d) Outdoor 8,16.}
\label{fig:CDF-S}
\end{figure*}

In this part of the experiment, we demonstrate the effect of downsampling in the time dimension with downsampling factors of 2, 4, 8, and 16. This includes downsampling the original IQ signal first and then obtaining the time-frequency representation through STFT (DD and AD), as well as directly downsampling the time-frequency representation in the time dimension (MPD and APD). Here, the pooling layer kernel size is $(1,D)$, with a stride of $(1,D)$.

As shown in Table \ref{table2}, in the indoor scenario, the method of downsampling the time-frequency representation generally performs better across different downsampling factors than the method of downsampling the IQ signal first. Specifically, with a downsampling factor of 2, the MDE values for DD and AD are 0.0518 and 0.0412, respectively, while the MDE values for MPD and APD are 0.0989 and 0.0494, indicating that direct downsampling of the IQ signal yields better results. However, as the downsampling factor increases, MPD and APD outperform the IQ signal downsampling methods in localization performance. Notably, with a downsampling factor of 16, MPD achieves MDE and STD values of 0.0256 and 0.2329, respectively, with localization accuracy even higher than without downsampling. In contrast, for the IQ signal downsampling methods, DD performs better, with MDE and STD values of 0.2475 and 0.9767, respectively, demonstrating that MPD offers superior localization performance and stability.

In the outdoor scenario, MPD consistently outperforms the other methods across all downsampling factors, with the performance gap widening as the downsampling factor increases. Specifically, for downsampling factors of 2, 4, 8, and 16, MPD achieves MDE values of 0.7794, 0.7628, 0.8143, and 0.7148, respectively. Among the other downsampling methods, APD performs best, with MDE values of 0.9517, 0.9257, 0.8488, and 1.1047 for downsampling factors of 2, 4, 8, and 16.

The CDF curves for the indoor scenario are shown in Fig. \ref{fig:indoor-s-2:ab} and Fig. \ref{fig:indoor-s-8:b}. MDP-4, MDP-8, and MDP-16 exhibit higher initial values and faster convergence speeds, with only DD-8 among the other methods showing comparable initial values and convergence rates. The CDF curve for the outdoor scenario in Fig. \ref{fig:outdoor-s-2:c} and Fig. \ref{fig:outdoor-s-8:d} further demonstrates that MDP performs better in terms of localization performance under downsampling factors of 2, 4, 8, and 16 in the outdoor environment.

Overall, MDP demonstrates better performance in both indoor and outdoor scenarios. This is because, compared to downsampling the IQ signals first (DD and AD), MDP preserves the most significant frequency components, thereby effectively maintaining the integrity of frequency information. In contrast, directly downsampling the IQ signal inevitably leads to the loss of some frequency components, resulting in a decrease in time-frequency resolution. Compared to APD, which also performs downsampling in the time dimension of the time-frequency representation, MDP is more effective because APD averages the values within the pooling window, which weakens the stronger features and reduces the clarity of the feature map and the distinguishability of the signal. In summary, MDP is more effective in retaining strong features during downsampling, leading to improved localization performance.

\subsection{Performance of Frequency Dimension Downsampling}
% \subsubsection{Effect of Sample Bandwidth}
\begin{table}[t]
\caption{Test Results of Different Downsampling Factors and Methods }
\centering
\renewcommand\arraystretch{1.0}
\setlength{\tabcolsep}{1.75mm}
\begin{tabular}{|c|c|cc|cc|}
\hline
\multirow{2}{*}{\begin{tabular}[c]{@{}c@{}}Downsampling\\   factor\end{tabular}} & \multirow{2}{*}{\begin{tabular}[c]{@{}c@{}}Downsampling\\    method\end{tabular}} & \multicolumn{2}{c|}{Indoor}          & \multicolumn{2}{c|}{Outdoor}         \\ \cline{3-6} 
                                                                                     &                                                                                      & \multicolumn{1}{c|}{MDE}    & STD    & \multicolumn{1}{c|}{MDE}    & STD    \\ \hline
\multirow{3}{*}{2}                                                                   & MPD                                                                                  & \multicolumn{1}{c|}{0.0164} & 0.1807 & \multicolumn{1}{c|}{0.6866} & 2.0763 \\ \cline{2-6} 
                                                                                     & APD                                                                                  & \multicolumn{1}{c|}{0.0273} & 0.2884 & \multicolumn{1}{c|}{0.7244} & 2.0577 \\ \cline{2-6} 
                                                                                     & AMD                                                                                  & \multicolumn{1}{c|}{0.0011} & 0.0477 & \multicolumn{1}{c|}{0.2330} & 1.5162 \\ \hline
\multirow{3}{*}{4}                                                                   & MPD                                                                                  & \multicolumn{1}{c|}{0.0446} & 0.3207 & \multicolumn{1}{c|}{0.7203} & 2.2110 \\ \cline{2-6} 
                                                                                     & APD                                                                                  & \multicolumn{1}{c|}{0.0626} & 0.4661 & \multicolumn{1}{c|}{0.7773} & 2.1054 \\ \cline{2-6} 
                                                                                     & AMD                                                                                  & \multicolumn{1}{c|}{0.0012} & 0.0508 & \multicolumn{1}{c|}{0.2485} & 1.6450 \\ \hline
\multirow{3}{*}{8}                                                                   & MPD                                                                                  & \multicolumn{1}{c|}{0.1776} & 0.7680 & \multicolumn{1}{c|}{0.6636} & 1.9526 \\ \cline{2-6} 
                                                                                     & APD                                                                                  & \multicolumn{1}{c|}{0.4987} & 1.2533 & \multicolumn{1}{c|}{2.0719} & 3.3795 \\ \cline{2-6} 
                                                                                     & AMD                                                                                  & \multicolumn{1}{c|}{0.0037} & 0.0628 & \multicolumn{1}{c|}{0.2728} & 1.6481 \\ \hline
\multirow{3}{*}{16}                                                                  & MPD                                                                                  & \multicolumn{1}{c|}{0.2289} & 0.7561 & \multicolumn{1}{c|}{0.9057} & 2.3991 \\ \cline{2-6} 
                                                                                     & APD                                                                                  & \multicolumn{1}{c|}{0.8799} & 1.3940 & \multicolumn{1}{c|}{2.5613} & 3.2079 \\ \cline{2-6} 
                                                                                     & AMD                                                                                  & \multicolumn{1}{c|}{0.0053} & 0.1160 & \multicolumn{1}{c|}{0.4416} & 2.5133 \\ \hline
\end{tabular}
\label{table3}
\end{table}

\begin{figure*}[t]
\centering
\subfigure[]{\label{fig:indoor-f-2:a}
\includegraphics[width=0.2\linewidth, height=3.5cm]{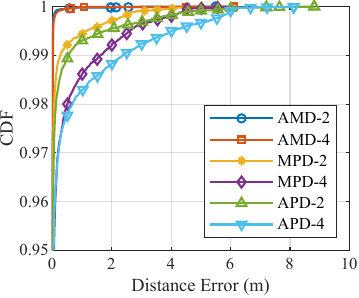}}
\hspace{0.02\linewidth}
\subfigure[]{\label{fig:indoor-f-8:b}
\includegraphics[width=0.2\linewidth, height=3.5cm]{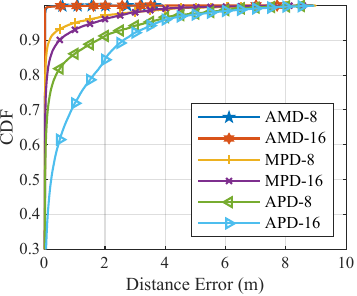}}
\hspace{0.02\linewidth}
\subfigure[]{\label{fig:outdoor-f-2:c}
\includegraphics[width=0.2\linewidth, height=3.5cm]{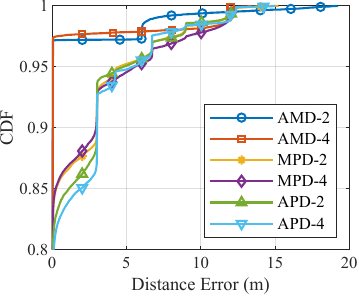}}
\hspace{0.02\linewidth}
\subfigure[]{\label{fig:outdoor-f-8:d}
\includegraphics[width=0.2\linewidth, height=3.5cm]{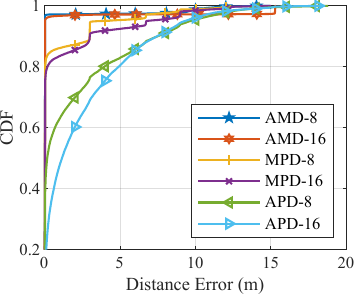}}
\hspace{0.02\linewidth}
% \subfigure[]{\label{fig:indoor:ca}
% \includegraphics[width=0.32\linewidth]{figs/1012.eps}}   
\centering
\caption{The Impact of Different Downsampling Factors and Methods on Localization Performance. (a) Indoor 2,4, (b) Indoor 8,16, (c) Outdoor 2,4, (d) Outdoor 8,16.}
\label{fig:CDF-F}
\end{figure*}

In the previous experiments, we have demonstrated the performance of downsampling in the time dimension. In this section, we will analyze the performance of downsampling the time-frequency representation in the frequency dimension. For this analysis, the pooling layer kernel size for MPD and APD is set to $(D,1)$, with a stride of $(D,1)$).

As shown in Table. \ref{table3}, in the indoor scenario, for downsampling factors of 2, 4, 8, and 16, the MDE values for AMD are 0.0011, 0.0012, 0.0037, and 0.0053, respectively, indicating that the AMD method achieves superior positioning performance. In contrast, MPD and APD perform worse across all downsampling factors, with the performance gap widening as the downsampling factor increases.

In the outdoor scenario, the AMD method again achieves the lowest MDE and STD values across all downsampling factors, indicating its superior localization accuracy in the outdoor environment compared to MPD and APD methods. Notably, under a downsampling factor of 16, the MDE for AMD is 0.4416, and the STD is 2.5133, which is significantly lower than the error values for MPD and APD, demonstrating the better localization performance of the AMD method.

The CDF curves for the indoor scenario are shown in Fig. \ref{fig:indoor-f-2:a} and Fig. \ref{fig:indoor-f-8:b}. Compared to the other two methods, the AMD method exhibits higher initial values and faster convergence rates in each CDF curve. Regardless of the downsampling factor for AMD, over 99\% of the samples achieve an MDE value below 0.05. In contrast, the other two methods struggle to achieve similar results across all downsampling factors. The CDF curves for the outdoor scenario are shown in Fig. \ref{fig:outdoor-f-2:c} and Fig. \ref{fig:outdoor-f-8:d}. In all downsampling factors, AMD achieves a distance error of less than 2 for 97\% of the samples, whereas the best-performing MPD-2 only achieves an MDE value lower than 2 for 87\% of the samples, showing a 10\% improvement in performance.

Overall, compared to MPD and APD, AMD consistently demonstrates superior localization performance across different downsampling factors. This is because MPD and APD lack the ability to adapt to different inputs, potentially losing key information or neglecting subtle but important features. On the other hand, AMD can adaptively select the frequency rows most relevant to the positioning task based on the attention mechanism. This more intelligent downsampling method allows for better retention of important local information, preventing the loss of critical features. As a result, DS-Pnet can effectively learn to improve localization performance.

When compared to downsampling in the time dimension, downsampling in the frequency dimension proves to be more effective. Specifically, in the time dimension, methods such as max pooling and average pooling perform well under moderate downsampling factors, but they are not as effective as the attention-based AMD method used in frequency dimension downsampling, especially at higher downsampling factors, where AMD maintains excellent localization performance while reducing computational complexity.

% \begin{figure}
% \centering
% \subfigure[]{
% \label{fig:different_bandwidth:a}
% \begin{minipage}[b]{0.4\textwidth}
% \includegraphics[width=1\textwidth]{figs/shiyan1_in_new.eps}
% \end{minipage}
% }
% \subfigure[]{
% \label{fig:different_bandwidth:b}
% \begin{minipage}[b]{0.4\textwidth}
% \includegraphics[width=1\textwidth]{figs/shiyan1_out_new.eps}
% \end{minipage}}
% \caption{The influence of different bandwidths on the positioning performance. (a) Indoor scenario, (b) outdoor scenario. }
% \label{fig:different_bandwidth}
% \end{figure}

% We use the data collected on Day1 in both indoor and outdoor scenarios. The sample length is fixed at 4,096. Datasets with different bandwidths are used for experiment, each trained separately. It's worth noting that in the case of a 320kHz bandwidth, the total number of samples is less than 4,000. Consequently, we use 2,000 samples for training and 1,000 samples for testing. The experimental results are presented in Table \ref{table2} and Fig. \ref{fig:different_bandwidth}. 
% \usepackage{multirow}

\begin{table}[]
\caption{Comparison of Different Downsampling Factors }
\centering
\renewcommand\arraystretch{1.0}
\setlength{\tabcolsep}{1.75mm}
\begin{tabular}{|c|c|c|cc|}
\hline
\multirow{2}{*}{Model} & \multirow{2}{*}{Parameters (M)} & \multirow{2}{*}{FLOPs (M)} & \multicolumn{2}{c|}{MDE}              \\ \cline{4-5} 
                       &                                &                           & \multicolumn{1}{c|}{Indoor} & Outdoor \\ \hline
ResNeXt                & 14.88                          & 17026.44                  & \multicolumn{1}{c|}{0.0531} & 0.9240  \\ \hline
MobileViT-XS           & 1.94                           & 1010.90                   & \multicolumn{1}{c|}{0.0503} & 0.7787  \\ \hline
MobileViT-XS-2         & 1.94                           & 503.42                    & \multicolumn{1}{c|}{0.0011} & 0.2330  \\ \hline
MobileViT-XS-4         & 1.94                           & 251.2                     & \multicolumn{1}{c|}{0.0012} & 0.2485  \\ \hline
MobileViT-XS-8         & 1.94                           & 130.5                     & \multicolumn{1}{c|}{0.0037} & 0.2728  \\ \hline
MobileViT-XS-16        & 1.94                           & 80.24                     & \multicolumn{1}{c|}{0.0053} & 0.4416  \\ \hline
\end{tabular}
\label{table4}
\end{table}
\subsection{Complexity Analysis}
In practical applications, the complexity of the model is a key factor in handling large-scale data and achieving real-time performance. %Complexity analysis aims to assess the efficiency of the model during both training and inference, focusing on time complexity and space complexity. 
We quantify computational complexity using floating-point operations (FLOPs) and measure space complexity using the number of trainable parameters (Params).

To compare the complexity of MobileViT-XS with the improved ResNeXt network, we conduct experiments with an input sample length of 16,384. After applying the STFT transformation, the time-frequency representation has a size of 256×257. The complexity results are shown in Table \ref{table4}. As indicated in the table, ResNeXt has 14.88 million parameters, whereas MobileViT-XS has only 1.94 million parameters, reducing the parameter count by approximately 87\%. This shows that ResNeXt is more complex and requires more memory for storing weight parameters. ResNeXt’s FLOPs are 17,026.44 million, significantly higher than MobileViT-XS’s 1,010.90 million, indicating that ResNeXt has higher computational demands and is more suitable for environments with abundant computational resources.

For a fixed neural network structure, the space complexity remains constant. Downsampling primarily serves to compress the input data dimensions, effectively reducing computational complexity. With the same downsampling factor, the compression effect on the input data dimensions is roughly consistent. Since the attention-based downsampling method performs better in the DS-Pnet framework, it is employed here. The original input data dimension is 256×257, and Table \ref{table4} details the time and space complexities and the corresponding MDE values for MobileViT-XS under various downsampling factors.

The results indicate that as the downsampling factor increases, both the MDE values and time complexity decrease, with the reduction in FLOPs being approximately proportional to the downsampling factor. Specifically, at a downsampling factor of 16, FLOPs are reduced by about $93\%$ compared to non-downsampling and by $99.5\%$ compared to FM-Pnet. When the downsampling factor is set to 2, the MDE values for indoor and outdoor scenarios are 0.0011 and 0.2330, respectively, reflecting significant improvements in positioning performance. With a downsampling factor of 16, the MDE values for indoor and outdoor scenarios are slightly higher at 0.0053 and 0.4416, but still better than the case without downsampling.

These results suggest that a downsampling factor of 16 achieves an optimal balance between computational efficiency and localization performance. Even at this high compression level, the model maintains excellent localization accuracy while significantly reducing the time and space complexity of the neural network.

\section{Conclusion}
\label{sec5}
%In this paper, we have proposed an FM-based positioning method called FM-Pnet, which is implemented using deep learning techniques. This method enables positioning by leveraging the time-frequency representations of FM signals. In order to validate the performance of FM-Pnet, we have conducted extensive experiments in both indoor and outdoor scenarios. The experimental results have demonstrated that our proposed FM-Pnet outperforms three existing RSS-based positioning methods in terms of both positioning accuracy and stability. The performance of FM-Pnet can be further enhanced in the frequency domain or in the time domain by increasing the bandwidth of the collected signals or extending the length of each sample. Additionally, through injecting noise in the training process or enriching training samples with cross-date signals, the generalization of FM-Pnet over a large time span can be improved. Therefore, the proposed FM-Pnet is a competitive positioning method for both indoor and outdoor scenarios, holding promising prospects for future applications. Future research could focus on exploring additional features inherent in FM signals that contribute to positioning accuracy, while also investigating lightweight deep learning neural networks to minimize complexity and enhance scalability for real-world deployment.
In conclusion, this paper proposes DS-Pnet, a novel framework for FM signal-based positioning that effectively balances localization accuracy and computational efficiency. By integrating lightweight deep learning models with innovative downsampling techniques, including IQ signal downsampling and time-frequency downsampling, DS-Pnet achieves significant reductions in computational complexity while maintaining robust positioning performance. Experimental results demonstrate that DS-Pnet outperforms traditional FM-based methods in both indoor and outdoor scenarios, achieving a parameter reduction of approximately 87\% and a FLOP reduction of up to 99.5\% compared to FM-Pnet, without compromising accuracy. These findings highlight the potential of FM signals for scalable and precise positioning solutions, particularly in resource-constrained environments. Future research will focus on further optimizing the framework for real-time applications and exploring its adaptability to other opportunistic signal sources, extending its versatility and practical impact.
 
\bibliographystyle{IEEEtran}%文件列表样式设定
\small
\bibliography{ref}%引用文件名

 \begin{IEEEbiography}[{\includegraphics[width=1in,height=1.25in,clip,keepaspectratio]{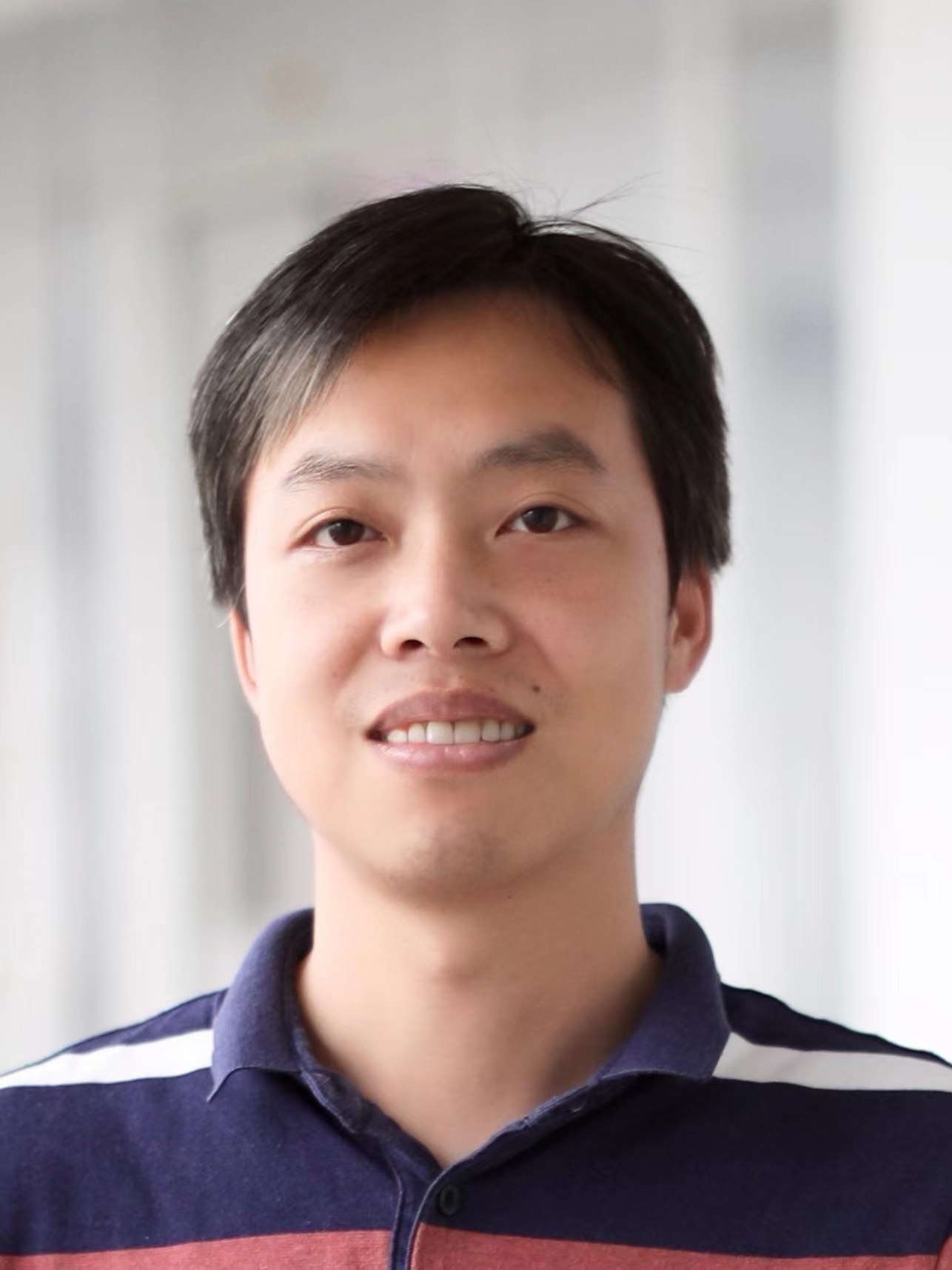}}]{Shilian Zheng}
 received the B.S. degree in telecommunication engineering and the M.S. degree in signal and information processing from Hangzhou Dianzi University, Hangzhou, China, in 2005 and 2008, respectively, and the Ph.D. degree in communication and information system from Xidian University, Xi'an, China, in 2014.

 He is currently a Researcher with National Key Laboratory of Electromagentic Space Security, Jiaxing, China, and a Doctoral Supervisor at Hangzhou Dianzi University, Hangzhou, China. His research interests include cognitive radio, deep learning-based radio signal processing, and electromagnetic space security.
 \end{IEEEbiography}

\begin{IEEEbiography}[{\includegraphics[width=1in,height=1.25in,clip,keepaspectratio]{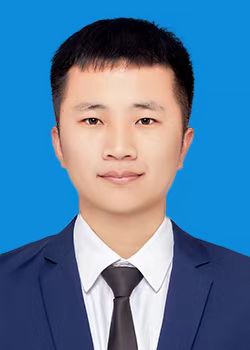}}]{Xinjiang Qiu}
 received the B.S. degree in Changchun University of Science and Technology, Changchun, China, in 2023. He is currently pursuing the M.S. degree with the School of Communications Engineering, Hangzhou Dianzi University, Hangzhou. His current research interests include deep learning and navigation via signals of opportunity.
\end{IEEEbiography}

% \begin{IEEEbiography}[{\includegraphics[width=1in,height=1.25in,clip,keepaspectratio]{photographs/Jiacheng Hu.jpg}}]{Jiacheng Hu}
% received the B.S. degree in Sichuan Normal University, Chendu, China, in 2022. He is currently pursuing the M.S. degree with the School of Communications Engineering, Hangzhou Dianzi University, Hangzhou. His current research interests include deep learning and navigation via signals of opportunity.
% \end{IEEEbiography}

\begin{IEEEbiography}[{\includegraphics[width=1in,height=1.25in,clip,keepaspectratio]{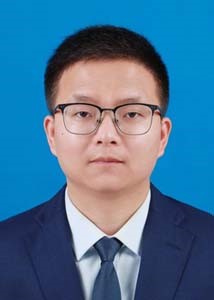}}]{Luxin Zhang}
received the M.S. degree in control science and engineering from Zhejiang University of Technology, Hangzhou, China, in 2021. 

He is currently an Assistant Engineer with National Key Laboratory of Electromagentic Space Security, Jiaxing, China. His research interests include cognitive radio, radio signal processing and learning-based radio signal recognition.
\end{IEEEbiography}

\begin{IEEEbiography}[{\includegraphics[width=1in,height=1.25in,clip,keepaspectratio]{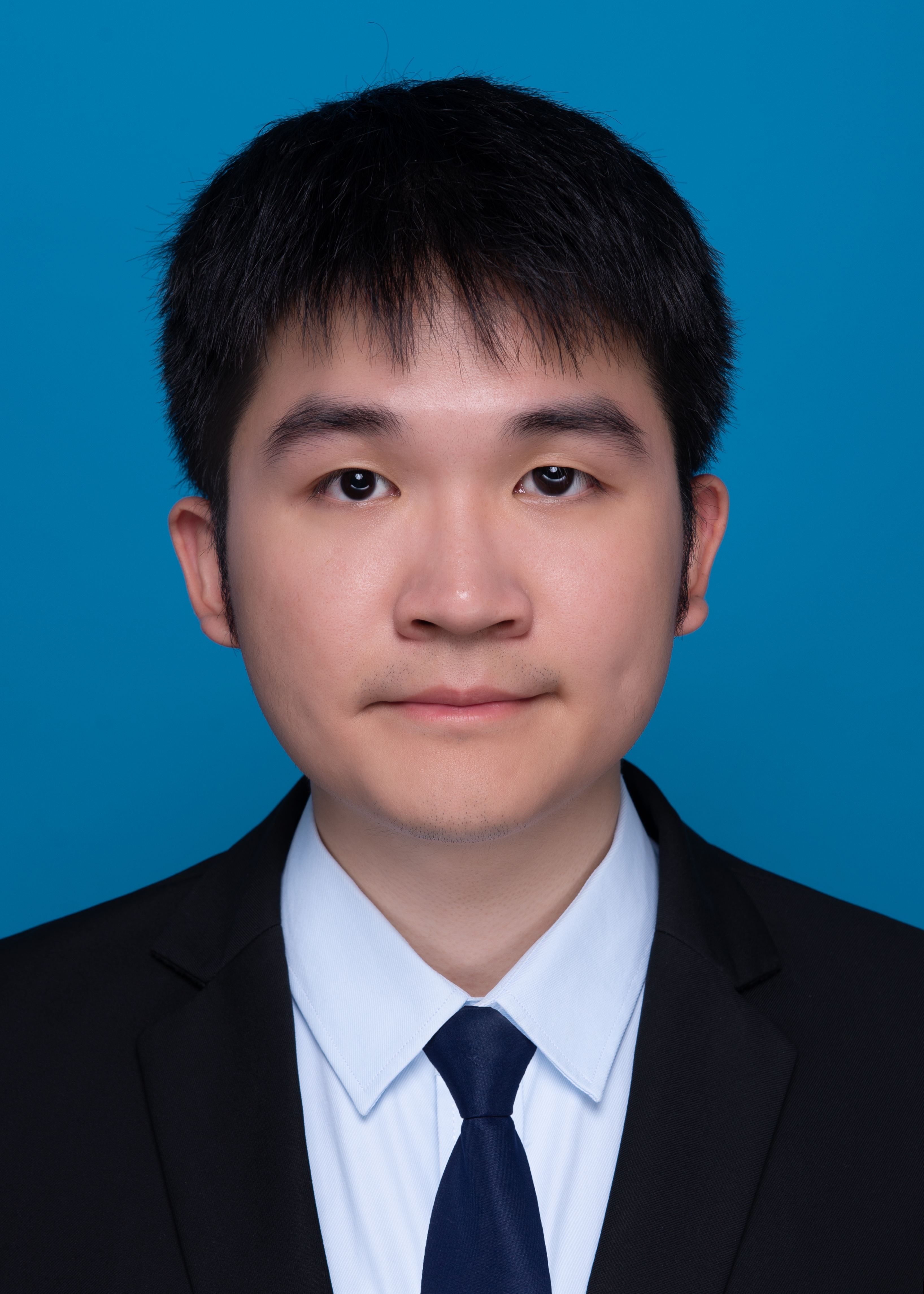}}]{Quan Lin}
 received the B.S. degree in Shaoxing University, Shaoxing, China, in 2023. He is currently pursuing the M.S. degree with the School of Communications Engineering, Hangzhou Dianzi University, Hangzhou. His current research interests include deep learning and navigation via signals of opportunity.
\end{IEEEbiography}

\begin{IEEEbiography}[{\includegraphics[width=1in,height=1.25in,clip,keepaspectratio]{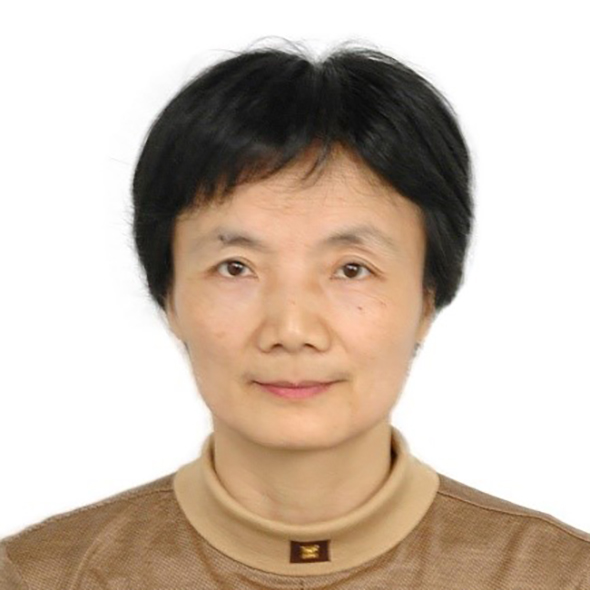}}]{Zhijin Zhao}
received the M.S. and Ph.D. degrees from from Xidian University, Xi’an, China, in 1984 and 2009, respectively. In 1993 and 2003, as a Visiting Scholar, she studied adaptive signal processing and blind signal processing in Darmstadt University of Technology and University of Erlangen-Nuremberg respectively. She is currently a Professor with the School of Communication Engineering, Hangzhou Dianzi University, Hangzhou, China. Her research interests include communication signal processing, cognitive radio technology, intelligent signal processing, and other aspects of research. She once served as President of the School of Communication Engineering, Hangzhou Dianzi University, as well as a Senior Member of China Electronics Society, a member of National Signal Processing Society.
\end{IEEEbiography}

\begin{IEEEbiography}[{\includegraphics[width=1in,height=1.25in,clip,keepaspectratio]{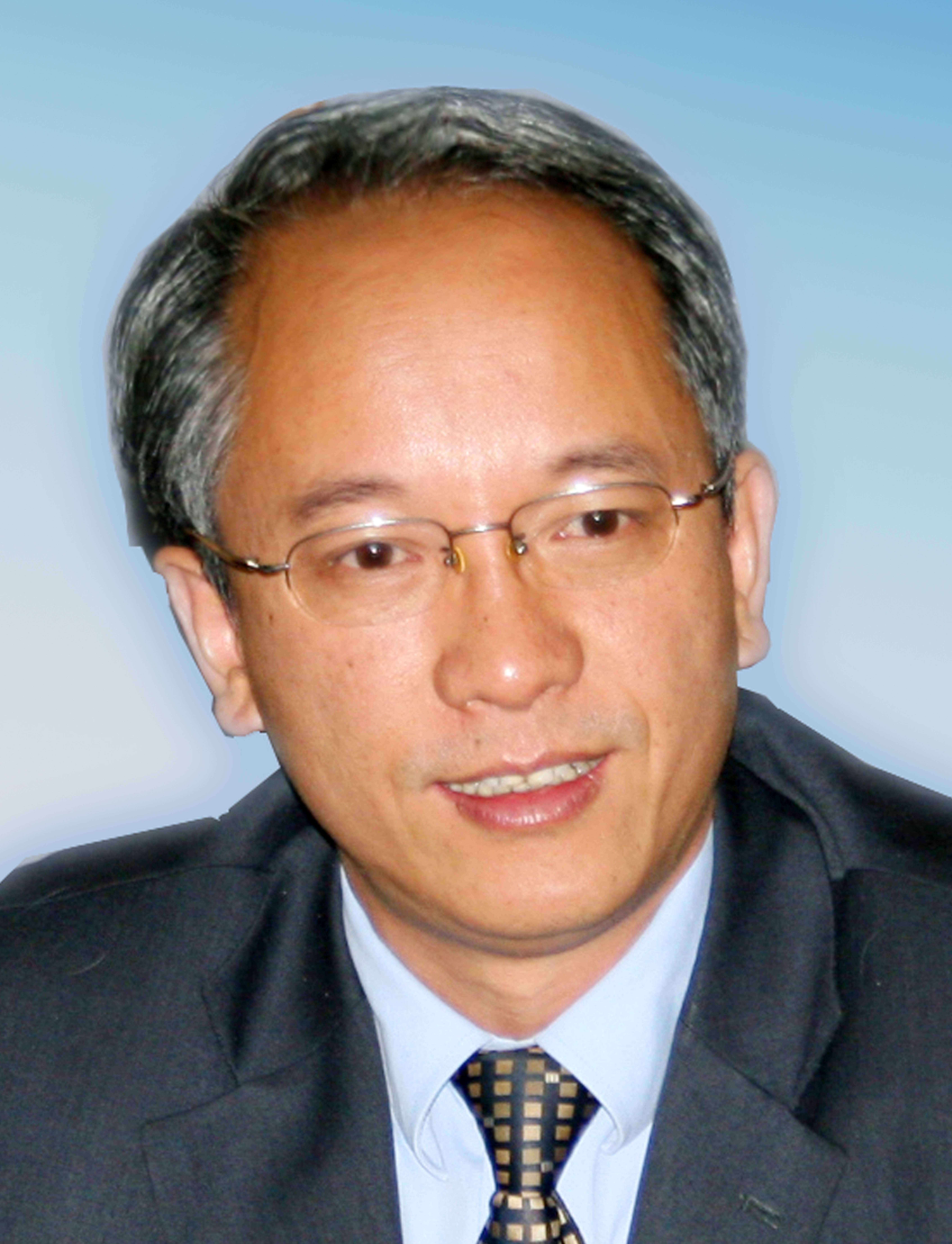}}]{Xiaoniu Yang}
is currently a chief scientist with National Key Laboratory of Electromagentic Space Security, Jiaxing, China. He is also an Academician of Chinese Academy of Engineering and a Fellow of the Chinese Institute of Electronics. He published the first software radio book in China (X. Yang, C. Lou, and J. Xu, Software Radio Principles and Applications, Publishing House of Electronics Industry, 2001 (in Chinese)). He holds more than 40 patents. His current research interests are  big data for radio signals and deep learning-based signal processing.
\end{IEEEbiography}

\end{document}